\documentclass{article}
\usepackage{graphicx}
\usepackage{pgfplots}
\usepackage{subfigure}
\usepackage{caption}
\usepackage[justification=centering]{caption}

\usepackage{amsmath}
\usepackage{amssymb}
\usepackage{listings}
\usetikzlibrary{arrows}
\usepackage{tikz}
\usepackage{float}
\usepackage[linesnumbered,ruled]{algorithm2e}
\usetikzlibrary{shapes.geometric, arrows}
\usepackage[utf8]{inputenc}
\usepackage{enumitem}
\usepackage[top=2.0cm, bottom=2.0cm, left=2.4cm, right=2.9cm]{geometry}
\usepackage{multicol}

\newtheorem{theorem}{Theorem}
\newtheorem{defn}{Definition}
\usepackage[titletoc,title]{appendix}
\usepackage{array}

\newcolumntype{P}[1]{>{\centering\arraybackslash}p{#1}}
\newcolumntype{M}[1]{>{\centering\arraybackslash}m{#1}}

\usetikzlibrary{arrows, chains, calc, positioning, decorations.markings, shadows, shapes.arrows}
\usepackage{authblk}

\usetikzlibrary{shapes}

\usepackage[superscript,biblabel]{cite}

\begin{document}

	\title{\textbf{The Effects of El Ni\~no on the Global Weather and Climate}}
	\author[1]{\textbf{Marat Akhmet}\thanks{Corresponding Author Tel.: +90 312 210 5355, Fax: +90 312 210 2972, E-mail: marat@metu.edu.tr}$^{,}$}
	\author[2]{\textbf{Mehmet Onur Fen}}
	\author[1]{\textbf{Ejaily Milad Alejaily}}
	\affil[1]{\textbf{Department of Mathematics, Middle East Technical University, 06800 Ankara, Turkey}}
	\affil[2]{\textbf{Department of Mathematics, TED University, 06420 Ankara, Turkey}}

	\date{}
	\maketitle
	
\begin{abstract}
\noindent This paper studies the chaotic behavior of hydrosphere and its influence on global weather and climate. We give mathematical arguments for the sea surface temperature (SST) to be unpredictable over the global ocean. The impact of SST variability on global climate is clear during global climate patterns, which involve large-scale ocean-atmosphere fluctuations similar to the El Ni\~no-Southern Oscillation (ENSO). Sensitivity (unpredictability) is the core ingredient of chaos. Several researches suggested that the ENSO might be chaotic. It was Vallis [\citen{Vallis}] who revealed unpredictability of ENSO by reducing his model to the Lorenz equations. Interactions of ENSO and other global climate patterns may transmit chaos. We discuss the unpredictability as a global phenomenon through extension of chaos ``horizontally'' and ``vertically'' in coupled Vallis ENSO models, Lorenz systems, and advection equations by using theoretical as well as numerical analyses. To perform theoretical research, we apply our recent results on replication of chaos and unpredictable solutions of differential equations, while for numerical analysis, we combine results on unpredictable solutions with numerical analysis of chaos in the advection equation.

\vspace{.2cm}

\end{abstract}

\vspace{.5cm}	
	
\noindent \textbf{The famous Lorenz equations give birth to the weather related observations. One of them is the unpredictability of weather in long period of time, which is a meteorological concept, and another one is that small changes of the climate and even weather at present may cause catastrophes for the human life in future. Issuing from this, we have taken into account the following three features of the Lorenz system, to emphasize the actuality of the present study. Firstly, it is a regional model. Secondly, for some values of its parameters the equations are non-chaotic. Finally, the model is of the atmosphere, but not of the hydrosphere. Therefore, one has to make additional investigations to reveal that the unpredictability of weather is a {\it global} phenomenon, and climatic catastrophes can be caused by physical processes at {\it any point} on the surface of the globe. The present paper is concerned with all of the three factors issuing from the ocean surface dynamics of ENSO type, and results of our former research.}

\section{Introduction and Preliminaries}
	
\subsection{Unpredictability of Weather and Deterministic Chaos}	
	
Global climate change has gained the attention of scientists and policymakers. The reason for that lies in its remarkable impact on human life on the Earth \cite{Roy}. Climate change affects and controls many social, economic and political human activities. It was an essential motive of human migration throughout history.
	
Weather is defined by the condition of the atmosphere at a specific place and time measured in terms of temperature, humidity, air pressure, wind, and precipitation, whereas climate can be viewed as the average of weather of a large area over a long period of time \cite{Petersen}. Some definitions of climate expand to include the conditions of not only the atmosphere, but also the rest components of the climate system: hydrosphere, cryosphere, lithosphere, biosphere and according to Vernadsky no{\"o}sphere \cite{Eppelbaum}.
	
During the last few decades, big efforts have been made to develop weather and climate change forecasting models. Due to the chaotic nature of weather, the forecasting range of weather prediction models is limited to only a few days. Climate models are more complicated than ordinary weather forecasting models, since they need to include additional factors of climate system that are not important in the weather forecast \cite{Schmandt}. Understanding the concepts of chaos is an important step toward better comprehension of the natural variability of the climate system on different  time scales. This involves determining what the reasons and sources that stand behind of presence of chaos in weather and climate models. Any progress made in this path will be helpful to adjust the conception of climate change and find solutions for climate control. 
	
Chaos can be defined as aperiodic long-term behavior in a deterministic system that exhibits sensitive dependence on initial conditions \cite{Strogatz}. Predictability consists of constructing a relationship between cause and effect by which we can predict and estimate the future behavior of a physical property. Unpredictability means the failure of such empirical or theoretical relationships to predict due to consisting of noise term(s), mathematical nature of the relationships or intrinsic irregularity of the physical property itself. Mathematically unpredictability is considered as a result of the sensitive dependence on initial conditions, which is an essential feature of Devaney chaos \cite{Devaney}. Recently, it is theoretically proved that a special kind of Poisson stable trajectory, called an unpredictable trajectory, gives rise to the existence of Poincar\'{e} chaos \cite{AkhmetExistence,AkhmetPoincare,AkhmetUnpredictable}.

Unpredictability in the dynamics of weather forecast models was firstly observed by E. N. Lorenz. He developed a heat convection model consisting of twelve equations describing the relationship between weather variables such as temperature and pressure. Lorenz surprisingly found that his system was extremely sensitive to initial conditions. Later, in his famous paper [\citen{Lorenz}], he simplified another heat convection model to a three-equation model that has the same sensitivity property \cite{Robinson}. This model is defined by the following nonlinear system of ordinary differential equations:
	\begin{equation} \label{LorenzSystem}
	\begin{split}
	& \frac{dx}{dt} =-\sigma\,x + \sigma y,\\
	& \frac{dy}{dt} = r\,x - xz-y,  \\	
	& \frac{dz}{dt} = xy - b \,z,
	\end{split}
	\end{equation}
where the variable $ x $ represents the velocity of the convection motion, the variable $ y $ is proportional to the temperature difference between the ascending and descending currents, and the variable $ z $ is proportional to the deviation of the vertical temperature profile from linearity, whereas the constants $ \sigma $, $ r $, and $ b $ are positive physical parameters. Model (\ref{LorenzSystem}) describes the thermal convection of a fluid heated from below between two layers. With certain values of these parameters, Lorenz system possesses intrinsic chaos and produces the so-called Lorenz butterfly attractor. 
	
The paper [\citen{AkhmetLorenz}] was concerned with the extension of chaos through Lorenz systems. It was demonstrated in [\citen{AkhmetLorenz}] that Lorenz systems can be unidirectionally coupled such that the drive system influences the response system, which is non-chaotic in the absence of driving, in such a way that the latter also possesses chaos. A possible connection of these results to the global weather dynamics was also provided in that study.

\subsection{Ocean-Atmosphere Interaction}	
	
Coupled ocean-atmosphere models are the most fundamental tool for understanding the natural processes that affect climate. These models have been widely applied to interpret and predict global climate phenomena such as ENSO \cite{Siedler}. In meteorology and climate science, SST is considered as a very important factor in ocean-atmosphere interaction, where it plays a basic role in determining the magnitude and direction of the current velocity, as well as the ocean surface wind speed. It is difficult to give a precise definition of SST due to the complexity of the heat transfer operations in the mixed layer of upper ocean. In general, however, it can be defined as the bulk temperature of the oceanic mixed layer with a depth varies from $ 1 \, m $ to $ 20 \, m $ depending on the measurement method used \cite{Barale}. The importance of SST stems from the fact that the world's oceans cover over $ 70 \, \% $ of the whole surface of the globe. This large contact area gives way to an active ocean-atmosphere interaction and sometimes becomes a fertile place for complex feedbacks between the ocean and atmosphere that drive an irregular climate change.
	
The most important example of the interactions and feedbacks between the ocean and the atmosphere is El Ni\~no and Southern Oscillation (ENSO) which is defined as a global coupled ocean-atmosphere phenomenon occurs irregularly in the Pacific Ocean about every $2$ to $7$ years \cite{Stuecker}. This phenomenon is accompanied by undesirable changes in weather across the tropical Pacific and losses in agricultural and fishing industries especially in South America. The El Ni\~no mechanism can be briefly summarized as follows: During normal conditions in the equatorial Pacific, trade winds blow from east to west driving the warm surface current in the same direction. As a consequence of this, warm water accumulates in the western Pacific around  southeast Asia and northern Australia. On the opposite side of the ocean around central and south America, the warm water, pushed to the west, is replaced by upwelling cold deep water. During El Ni\~no conditions, the trade winds are much weaker than normal. Because of this and due to SST difference, warm water flows back towards the western Pacific. This situation involves large changes in air pressure and rainfall patterns in the tropical Pacific. The cool phase of this phenomenon is called La Nina, which is an intensification of the normal situation. The term ``Southern Oscillation'' is usually used to refer to the difference of the sea-level pressure (SLP) between Tahiti and Darwin, Australia. Bjerknes \cite{Bjerknes} conclude that El Ni\~no and the Southern Oscillation are merely two different results of the same phenomenon. These phases of the phenomenon are scientifically called El Ni\~no Southern Oscillation or shortly ENSO. From the above mechanism we can note that the ENSO dynamics is a perfect example of self-excited oscillating systems.
	
The ramifications of El Ni\~no are not restricted to the Pacific basin alone, but have widespread effects which severely disrupt global weather patterns. In the last few decades scientists developed theories about the climatic engine which produced El Ni\~no, and they are trying to explain how that engine interact with the great machine of global climate. Although remarkable progress has been made in monitoring and forecasting the onset of El Ni\~no, it is still challenging to predict its intensity and the impact of the event on global weather. Study of ENSO is considered as a key to understanding climate change, it is a significant stride toward the meteorology's ultimate goal, ``accurate prediction and control of world weather''.
		
Besides the ENSO, there are several other atmospheric patterns that occur in different regions of the Earth. These phenomena are interacting in very complicated ways. Many researchers paid attention to the mutual influence of these phenomena and investigated if there is any co-occurrence relationship or interaction between them.  
	
The most similar atmosphere-ocean coupled phenomenon to ENSO is the Indian Ocean Dipole (IOD), which occur in the tropical Indian Ocean, and it is sometimes called the  Indian Ni\~no. IOD has normal (neutral), negative and positive phases. During neutral phase, Pacific warm water, driven by the Pacific trade winds, cross between south Asia and Australia and flow toward the Indian Ocean. Because of the westerly wind, the warm water accumulates in the eastern basin of Indian Ocean. In the negative IOD phase with the coincidence of strength of the westerly wind, warmer water concentrate near Indonesia and Australia, and cause a heavy rainfall weather in these regions and cooler SST and droughts in the opposite side of the Indian Ocean basin around the eastern coast of Africa. The positive phase is the reversal mode of the negative phase, i.e., what happened in the east side will happen in west side and vice versa.
		
From the above we can see that there is a symmetry between the IOD and ENSO mechanisms. Indeed, SST data shows that the Indian Ocean warming appears as a near mirror image of ENSO in the Pacific \cite{Chambers}. In addition, the IOD is likely to have a link with ENSO events, where a positive (negative) IOD often occurs during El Ni\~no (La Nina) \cite{Eamus, Yamagata}. Luo et al. \cite{Luo} investigated the ENSO-IOD interactions, and they suggest that IOD may significantly enhance ENSO and its onset forecast, and vice versa. Several other researchers like \cite{Behera, Roxy} studied the relationship and interaction between ENSO and IOD. It should be noted here that (as in all these studies) the SST considered as the major variable, indicator and index for these events.

Other important atmosphere-ocean coupled phenomena are briefly described in Table \ref{T1} (Appendix) and Figure \ref{Wmap} (Appendix) shows the places of occurrence of them. They have significant influences on weather and climate variability throughout the world. Similar to the relationship between ENSO and IOD, various studies show expected relationships between these phenomena and mutual effects on their predictability. These pattern modes have different degrees of effect on SST. In Table \ref{T1} (Appendix) we see that the patterns that remarkably influence the ocean temperature are indexed by SST, whereas those that are most correlated with air pressure, the main indexes of them are based on SLP. 
	
\subsection{El Ni\~no Chaos} \label{ElNinoChaos}
		
\begin{figure}[H]
\centering
\includegraphics[width=1.0\linewidth]{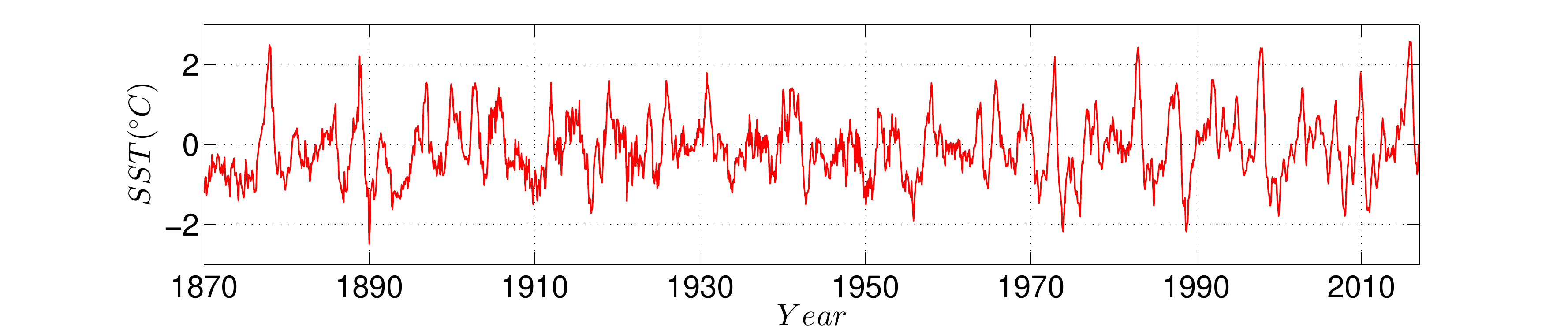}
\caption{Sea surface temperature anomalies of NINO3.4 region. The data utilized in the figure is from the Hadley Centre Sea-Ice and SST dataset HadISST1 \cite{Rayner}.}
\label{SST_Curve}	
\end{figure}

The SST behavior associated with ENSO indicates irregular fluctuations. The ENSO indicator NINO3.4 index, for example, is one of the most commonly used indices, where the SST anomaly averaged over the region bounded by $ 5^{\circ}$N--$5^{\circ}$S, $170^{\circ}$--$120^{\circ}$W \cite{Bunge}. Figure \ref{SST_Curve} shows the oscillatory behavior of SST in the NINO3.4 region. Data from the Hadley Centre Sea-Ice and SST dataset HadISST1 \cite{Rayner} is used to generate the figure. This behavior encourages many scientists to answer the question: Is ENSO a self-sustained chaotic oscillation or a damped one, requiring external stochastic forcing to be excited? \cite{Sheinbaum}. 
There are different hypotheses for the source of chaos in ENSO. According to Neelin and Latif \cite{Neelin}, deterministic chaos within the nonlinear dynamics of coupled system, uncoupled atmospheric weather noise and secular variation in the climatic state are the possible source of ENSO irregularity. Tiperman et al. \cite{Tziperman} concluded that the chaotic behavior of ENSO is caused by the irregular jumping of the ocean-atmosphere system among different nonlinear resonances. Several studies like \cite{Battisti, Penland} support this assumption and attributed the irregularity and the unpredictability of ENSO to influence of stochastic forcing generated by weather noise. Other studies like \cite{Zebiak, Munnich} infer that ENSO is intrinsically chaotic, which means that the irregularity and the loss of predictability are independent of the chaotic nature of weather.

Practically, investigating chaos in ENSO needs long time-series of data, which make the task quite difficult experimentally. Vallis \cite{Vallis}, developed a conceptual model of ENSO and suggested that the ENSO oscillation exhibits a chaotic behavior. Vallis used finite difference formulation to reduce two dimensional versions of advection and continuity equations to a set of ordinary differential equations. In addition, he assumed that the zonal current is driven by the surface wind, which is in turn proportional to the temperature difference across the ocean. The model is described by the set of equations	
\begin{equation} \label{VallisModel}
\begin{split}
\frac{du}{dt} & =\beta \,(T_e-T_w)-\lambda \,(u-u^*),\\
\frac{dT_w}{dt} & = \frac{u}{2\Delta x} \,(\bar{T}-T_e)-\alpha \,(T_w-T^*) ,  \\	
\frac{dT_e}{dt} & = \frac{u}{2\Delta x} \,(T_w-\bar{T})-\alpha \,(T_e-T^*),
\end{split}
\end{equation}
where $ u $ represents the zonal velocity,  $ T_w $ and $ T_e $ are the SST in the eastern and western ocean respectively, $ \bar{T} $ is the deep ocean temperature, $ T^* $ is the steady state temperature of ocean, $ u^* $ represents the effect of the mean trade winds, $ \Delta x $ is the width of the ocean basin, and $\alpha,$ $\beta$ and $\lambda$ are constants.
	
By nondimensionalizing Equations (\ref{VallisModel}) and forming the sum and difference of the two temperature equations, one can see that these equations have the same structure as the Lorenz system (\ref{LorenzSystem}). Vallis utilized the fact that the Lorenz system, with specific parameters, is intrinsically chaotic, and showed that a chaotic behavior of the sum and difference of the west and east SST can be obtained. 
	
ENSO, as mentioned above, occurs as a result of the interaction of the ocean and atmosphere. Therefore, modeling of ENSO would be a good instrument to research unpredictability not only in the atmosphere but also in the hydrosphere. Nevertheless, ENSO provides the arguments that unpredictability is also proper for sea water parameters which possibly can be reduced to a single one, the SST, if one excludes flow characteristics.
Vallis saved in the model only hydrosphere variables ignoring the variation of atmosphere parameters when he considers chaos problem. In our opinion, however, the model is appreciated as a pioneer one, and furthermore, it implies chaos presence in the Pacific ocean water. Hopefully, in the future, ENSO with both atmosphere and hydrosphere characteristics being variable will be modeled, but this time we focus on chaotic effects of ENSO by utilizing the Vallis model.
	
\subsection{SST Advection Equation}
	
The temporal and spatial evolution of the SST is governed by a first order quasi-linear partial differential equation, the advection equation. If we denote the SST by $ T $, the temperature advection equation of mixed layer of fixed depth can be written in the form \cite{Willebrand,Lucas}
\begin{equation} \label{AdvectionEq}
\frac{\partial T}{\partial t} + u \frac{\partial T}{\partial x} + v \frac{\partial T}{\partial y} + w \frac{\partial T}{\partial z}= f(t,x,y,z,T),
\end{equation}	
where $ u, v, w $ are the zonal, meridional and vertical components of current velocity, respectively. These velocities theoretically must satisfy the continuity equation
\begin{equation} \label{ContinuityEq}
\frac{\partial}{\partial x}(\rho u) + \frac{\partial}{\partial y}(\rho v) + \frac{\partial}{\partial z}(\rho w) = - \frac{\partial \rho}{\partial t},
\end{equation}	
where $ \rho $ is the seawater density.
		
The inhomogeneous (forcing) term $ f $ on the right-hand side of Equation (\ref{AdvectionEq}) consists of the shortwave flux, the evaporative heat flux, the combined long-wave back-radiation and sensible heat flux and heat flux due to vertical mixing \cite{Kessler}. This term can be described by \cite{Gent, Stevenson, Jochum}
\begin{equation} \label{ForceTermEq}
f \approx \frac{1}{h  \rho \, C_p} \, \frac{\partial q}{\partial z} + D,
\end{equation}	
where $ h $ is the mixed layer depth, $ C_p $ is the heat capacity of seawater, $ q $ is radiative and diffusive heat flux, and $ D $ is the thermal damping (the numerical diffusion operator). 
	
The spatial and temporal domain of Equation (\ref{AdvectionEq}) depend on the region and the nature of the phenomenon under study. For studying ENSO or IOD, for instance, there are various regions for monitoring SST. NINO3.4 is one of the most commonly used indices for ENSO. Dipole Mode Index (DMI) is usually used for IOD, and it depends on the difference in average SST anomalies between the western $50^{\circ}$E--$70^{\circ}$E, $10^{\circ}$N--$10^{\circ}$S and the eastern $90^{\circ}$E--$110^{\circ}$E, $0^{\circ}$--$10^{\circ}$S boxes \cite{Saji}. The mixed layer depth $ h $ varies with season and depends on the vertical heat flux through the upper layers of the ocean. The average of mixed layer depth is about $ 30 \, \text{m} $ \cite{Lukas}. Different studies of ocean-atmosphere coupled models considered different regions of various sizes. Zebiak and Cane \cite{Zebiak}, for example, developed a model of ENSO. They considered a rectangular model extending from $ 124^{\circ} $E to $ 80^{\circ} $W and $ 29^{\circ} $N to $ 29^{\circ} $S, with constant mixed layer depth of $ 50 \, \text{m} $ and 90 years simulation.
	
From the above we find that the domain of Equation (\ref{AdvectionEq}) depends on the purpose of the study. To study ENSO, for instance, we would cover a big region of pacific ocean basin, and if we choose the origin of coordinates to be at $160^{\circ}$E on the Equator, we can write the domain of (\ref{AdvectionEq}) as follows
\begin{equation*} \label{DomainMainEq}
t \geq 0, \quad 0\leq x \leq 9000 \, \text{km} , \quad -3000 \, \text{km} \leq y \leq 3000 \, \text{km}, \quad -100 \, \text{m} \leq z \leq 0.
\end{equation*}
	
The inhomogeneous term in Equation (\ref{AdvectionEq}), which includes mixing processes of heat transfer, plays the main role for chaotic dynamics. In addition to this term, a chaotic behavior in ocean current velocity terms may also produce an unpredictable behavior in SST. These causes of unpredictability are proved analytically and numerically by perturbing these terms by unpredictable functions. In this study we treat Equation (\ref{AdvectionEq}) mathematically without paying attention to the dimensions of the physical quantities. The important thing to us is the possibility of presence of chaos in this advection equation endogenously or be acquired from other equation or system. The advection equation, in addition to the Vallis model and the Lorenz system, will be used to demonstrate the extension of unpredictability ``horizontally'' through the global ocean and ``vertically'' between ocean and atmosphere.

\subsection{Unpredictable Functions and Chaos} \label{SubSec:UnpFuncChaos}
	
There are different types and definitions of chaos. Devaney \cite{Devaney} and Li-Yorke \cite{Yorke} chaos are the most frequently used types, which are characterized by transitivity, sensitivity, frequent separation and proximality. Another common type is the period-doubling cascade, which is a sort of route to chaos through local bifurcation \cite{Feigenbaum80,Scholl,SanderYorke11}. In the papers [\citen{AkhmetUnpredictable,AkhmetPoincare}], Poincar\'{e} chaos was developed by introducing the theory of unpredictable point and unpredictable function, which are built on the concepts of Poisson stable point and function. We define unpredictable point as follows. Let $(X, d)$ be a metric space and $ \pi : \mathbb{T} \times X \to X $ be a flow on X, where $ \mathbb{T} $ refer to either the set of real numbers or the set of integers. We assume that the triple $ ( \pi, X, d ) $ defines a dynamical system.
	
\begin{defn}  ([\citen{AkhmetUnpredictable}]) \label{UnpPointDiff}
A point $ p \in X  $ and the trajectory through it are unpredictable if there exist a positive number $ \epsilon $ (the unpredictability constant) and sequences $\left\{t_n\right\}$, $\left\{\tau_n\right\}$ both of which diverge to infinity such that $ \displaystyle{\lim_{n \to \infty}} \pi(t_n , p) = p $ and $ d[ \pi(t_n + \tau_n , p) , \pi(\tau_n , p)] \geq \epsilon $ for each $ n \in \mathbb{N} $.
\end{defn}
Definition \ref{UnpPointDiff} describes unpredictability as individual sensitivity for a motion, i.e., it is formulated for a single trajectory. The Poincar\'{e} chaos is distinguished by Poisson stable motions instead of periodic ones. Existence of infinitely many unpredictable Poisson stable trajectories that lie in a compact set meet all requirements of chaos. Based on this, chaos can be appeared in the dynamics on the quasi-minimal set which is the closure of a Poisson stable trajectory. Therefore, the Poincar\'{e} chaos is referred to as the dynamics on the quasi-minimal set of trajectory initiated from unpredictable point.
	
The definition of an unpredictable function is as follows.
	
\begin{defn} ([\citen{AkhmetUnpredSolnDE}]) \label{UnpFuncDiff}
A uniformly continuous and bounded function $ \varphi: \mathbb{R} \rightarrow \mathbb{R}^m $ is unpredictable if there exist positive numbers $\epsilon$, $\delta$ and sequences $\left\{t_n\right\}$, $\left\{\tau_n\right\}$ both of which diverge to infinity such that $ \| \varphi(t+t_n) - \varphi(t) \| \rightarrow 0 $ as $ n \rightarrow \infty $ uniformly on compact subsets of $ \mathbb{R} $, and $ \| \varphi(t+t_n) - \varphi(t) \| \geq  \epsilon $ for each $ t \in [ \tau_n - \delta , \tau_n +\delta ] $ and  $ n \in \mathbb{N} $.
\end{defn}
To determine unpredictable functions, we apply the uniform convergence topology on compact subsets of the real axis. This topology allows us to construct Bebutov dynamical system on the set of the bounded functions \cite{AkhmetExistence,Sell}. Consequently, the unpredictable functions imply presence of the Poincar\'{e} chaos.

\subsection{Global Weather and Climate}
	
The topic of weather and climate is one of the most profoundly important issues concerning the international community. It becomes very actual since the catastrophic phenomena such as global warming, hurricanes, droughts, and floods. This is why weather and climate are agenda of researches in physics, geography, meteorology, oceanography, hydrodynamics, aerodynamics and other fields. The problem is global, that is a  comprehensive model would include the interactions of all major climate system components, howsoever, for a specific aspect of the problem, a appropriate model combination can be considered \cite{Stocker}. In the second half of the last century, it was learned \cite{Lorenz} that the weather dynamics is irregular and sensitive to initial conditions. Thus the chaos was considered as a characteristic of weather which can not be ignored. Moreover, chaos can be controlled \cite{Ott,Pyragas}. These all make us optimistic that the researches of weather and climate considering chaos effect may be useful not only for the deep comprehension of their processes but also for control of them. In our research \cite{AkhmetReplication}, we have shown how a local control of chaos can be expanded globally. 
	
It is not wrong to say that in meteorological studies, chaos is considered as a severe limiting factor in the ability to predict weather events accurately \cite{Saravanan}. Beside this one can say that chaos is also a responsible factor for climate change if it is considered as a weather consequence. This is true, firstly, because of the weather unpredictability, since predictability can be considered as a useful feature of climate with respect to living conditions, and secondly, as the small weather change may cause a global climate change in time. Accordingly, it is possible to say that the control of weather, even a limited artificial one, bring us to a change of climate.	
	
The chaotic behavior has also been observed in several models of ENSO \cite{Neelin}. Presence of chaos in the dynamic of this climate event provides other evidence of the unpredictable nature of the global weather. Besides the Lorenz chaos of atmosphere, ``Vallis chaos'' takes place in the hydrosphere. Without exaggerating, we can say that chaos seems to be inherent at the essence of any deterministic climate model. Therefore, unpredictability can be globally widespread phenomenon through constructive interactions between the components of the climate system. 
	
To give a sketch how chaos is related globally to weather and climate, we will use, in the present research, information on dynamics of ENSO which will mainly utilize the Vallis model as will as the SST advection equation and the Lorenz equations. They will be properly coupled to have the global effect. It is apparent that, in the next research, the models will possibly be replaced by more developed ones, but our main idea is to demonstrate a feasible approach to the problem by constructing a special net of differential equations system. Obviously, one can consider the net as an instrument which can be subdued to an improvement by arranging new perturbation connections.
	
Proceeding from aforementioned remarks and as a part of the scientific work, we focus on one possible aspect of global weather and climate dynamics based on El Ni\~no phenomenon. To address this aim, we first review the Vallis model research for El Ni\~no in Subsection \ref{ElNinoChaos}, then, in Section \ref{UnpAdvEq} we analyze the presence of chaos in isolated models for the SST advection equation. In Section \ref{HorizontalExtension}, the extension of chaos in hydrosphere discussed through coupling of advection equation, the Vallis model and also mixing advection equation with the Vallis model. In paper \cite{AkhmetLorenz} we considered chaos as a global phenomenon in atmosphere, but it is clear that, to say about the globe weather one should take into account hydrosphere as well as the interaction processes between atmosphere and seas. For this reason Section \ref{VerticalExtension} is written where chaos extension from ocean to air and vice versa is discussed on the base of the Lorenz and Vallis models. So, finalizing the introduction we can conclude that the present paper is considered as an attempt to give a sketch of the global effects of chaos on weather and climate. This results are supposed to be useful for geographers, oceanographers, climate researchers and those mathematician who are looking for chaotic models and theoretical aspects of chaos researching. 
	
\section{Unpredictable Solution of the Advection Equation} \label{UnpAdvEq}
	
In this section we study the presence of Poincar\'{e} chaos in the dynamics of Equation (\ref{AdvectionEq}). We expect that the behavior of the solutions of (\ref{AdvectionEq}) depends on the function $ f $ and the current velocity components $ u, v, w $, which are used in the equation. From Equation (\ref{ForceTermEq}), we see that the forcing term $ f $ depends mainly on the heat fluxes between the sea surface and atmosphere which is governed by SST, air temperature and wind speed, as well as between layers of sea which is caused by sea temperature gradient and vertical (entrainment) velocity. Therefore, this forcing term can be a natural source of noise and irregularity. Ocean currents are mainly driven by wind forces, as well as temperature and salinity differences \cite{Coley}. Thence again we can deduce that the irregular fluctuations of wind may be reflected in the behavior of SST.
	
To demonstrate the role of the function $ f $ in the dynamics of Equation (\ref{AdvectionEq}), let us take into account the equation
\begin{eqnarray}  \label{Assump2Eq}
\frac{\partial T}{\partial t} + u \frac{\partial T}{\partial x} + v \frac{\partial T}{\partial y} + w \frac{\partial T}{\partial z}= -0.7 \, T+0.3 \, w_1 \, T+5 \, \sin(xt),
\end{eqnarray}
where the current velocity components are defined by $u= \sin(\frac{_x}{^{2}})+\sin(t)+3,$ $v= -0.02,$ and $w= -\frac{_1}{^{2}}  \cos(\frac{_x}{^{2}}) z.$

\begin{figure}[H]
		\centering
		\includegraphics[width=0.9\linewidth]{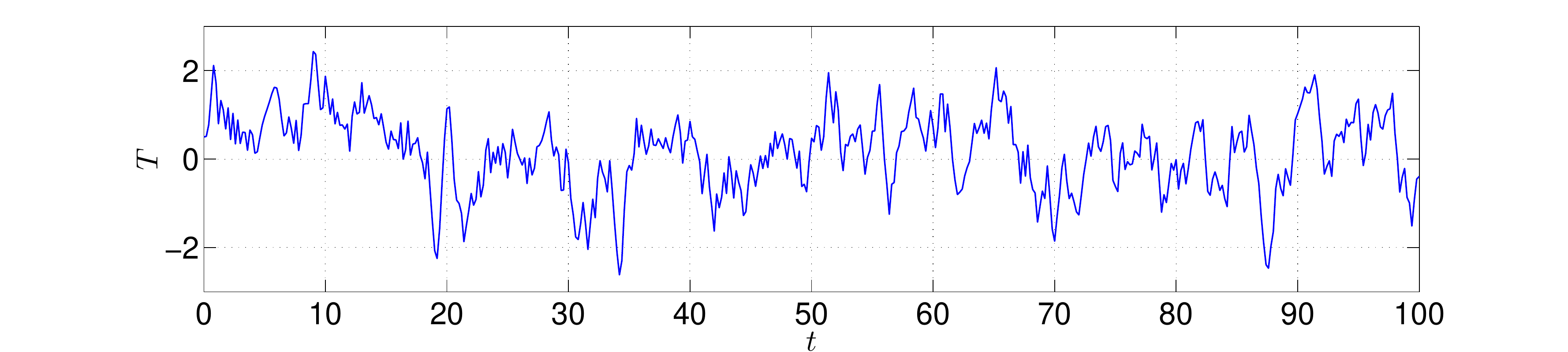}
		\caption{The solution of Equation (\ref{Assump2Eq}) with the initial condition $T(0, 0, 0, 0)= 0.5$. The figure shows that the forcing term $ f $ has a significant role in the dynamics of (\ref{AdvectionEq}).}
		\label{SSTforAssump2Eq}	
	\end{figure}

	Figure \ref{SSTforAssump2Eq} represents the solution of (\ref{Assump2Eq}) corresponding to the initial data $ T(0, 0, 0, 0)= 0.5 $. It is seen in Figure \ref{SSTforAssump2Eq} that the solution of Equation (\ref{Assump2Eq}) has an  irregular oscillating behavior, whereas in the absence of the term $ 5 \sin(xt) $ in the function $ f $,  the solution approaches the steady state. Even though the behavior of this numerical solution depends on the step size of the numerical scheme used, this situation leads us to consider that the forcing term $ f $ has a dominant role in the behavior of SST. \\

	To investigate the existence of an unpredictable solution in the dynamics of Equation (\ref{AdvectionEq}) theoretically, let us apply the method of characteristics. If we parametrize the characteristics by the variable $ t $ and suppose that the initial condition is given by $ T(t_0,x,y,z)=\Phi (x,y,z) $, where $  t_0 $ is the initial time, then we obtain the system
	
	\begin{equation} \label{ParamSysEq}
	\begin{split}
	& \frac{dx}{dt} = u(t,x,y,z,T) , \\
	& \frac{dy}{dt} = v(t,x,y,z,T) , \\	
	& \frac{dz}{dt} = w(t,x,y,z,T) , \\		
	& \frac{dT}{dt} = f(t,x,y,z,T) , 
	\end{split}
	\end{equation}
	with the initial conditions
	\begin{equation*} \label{InitialCondParamSysEq}
	x(t_0)=x_0, \quad  y(t_0)=y_0, \quad  z(t_0)=z_0, \quad  T(t_0,x_0,y_0,z_0)=\Phi (x_0,y_0,z_0).	
	\end{equation*}
	In system (\ref{ParamSysEq}), we assume that $ u, v, w $, and $ f $ are functions of $ x,y,z,t $, and $ T $, and they have the forms
	\begin{equation} \label{Velo&TfuncEq}
	\begin{split}
	& u=a_1 \, x+a_2 \, y+a_3 \, z+a_4 \, T+U(x, y, z, T),  \\
	& v=b_1 \, x+b_2 \, y+b_3 \, z+b_4 \, T+V(x, y, z, T),  \\
	& w=c_1 \, x+c_2 \, y+c_3 \, z+c_4 \, T+W(x, y, z, T),  \\
	& f=d_1 \, x+d_2 \, y+d_3 \, z+d_4 \, T+F(x, y, z, T),
	\end{split}
	\end{equation}
	where $ a_i $, $ b_i $, $ c_i $, $ d_i $, $i=1,2,3,4$, are real constants and the functions $ U, V, W, F $ are continuous in all their arguments. System (\ref{ParamSysEq}) can be expressed in the form
	\begin{equation} \label{SysODEEq}
	X'(t) = A  X(t) + Q(t), 	
	\end{equation}
	in which
	\begin{equation} \label{Eq1}
	X(t)=
	\begin{bmatrix}
	x \\
	y \\
	z \\
	T 
	\end{bmatrix}, \qquad 
	A= 	
	\begin{bmatrix}
	a_1 & a_2 & a_3 & a_4 \\
	b_1 & b_2 & b_3 & b_4 \\
	c_1 & c_2 & c_3 & c_4 \\
	d_1 & d_2 & d_3 & d_4	
	\end{bmatrix}, \qquad
	Q=
	\begin{bmatrix}
	U \\
	V \\
	W \\
	F 
	\end{bmatrix}. 
	\end{equation}
	
	The following theorem is needed to verify the existence of Poincar\'{e} chaos in the dynamics of Equation (\ref{AdvectionEq}).
	
	\begin{theorem} \label{Thm1} ([\citen{AkhmetPoincare}])
		Consider the system of ordinary differential equations
	\begin{equation} \label{SysODEThmEq}
	X'(t)= A \, X(t) + G(X(t))+H(t), 	
	\end{equation}
		where the $ n \times n $ constant matrix $ A $ has eigenvalues all with negative real parts, the function $ G: \mathbb{R}^n \rightarrow \mathbb{R}^n  $ is Lipschitzian with a sufficiently small Lipschitz constant, and $ H: \mathbb{R} \rightarrow \mathbb{R}^n $ is a uniformly continuous and bounded function. If the function $ H(t) $ is unpredictable, then system (\ref{SysODEThmEq}) possesses a unique uniformly exponentially stable unpredictable solution, which is uniformly continuous and bounded on the real axis. \\
		
	\end{theorem}

	In the remaining parts of the present section, we will discuss the unpredictability when SST is chaotified by external irregularity. For that purpose let us consider the logistic map
		\begin{equation} \label{logistic} 
		\eta_{j+1}= 3.91 \, \eta_j \, (1- \eta_j), \ j\in \mathbb Z.
		\end{equation}
	According to Theorem $4.1$ [\citen{AkhmetPoincare}], the map (\ref{logistic}) is Poincar\'{e} chaotic such that it possesses an unpredictable trajectory.

	 Next, we define a function $ \phi(t) $ by
		\begin{equation} \label{UnpFunInEq}
		\phi(t)=\int_{-\infty}^t e^{-2(t-s)} \gamma^*(s)\,ds,
		\end{equation}
where
	\begin{equation}\label{relayfunc}
	\gamma^*(t) =
	\begin{cases}
	1.5, & \zeta^*_{2j} < t \leq \zeta^*_{2j+1}, \quad j \in \mathbb{Z}, \\
	0.3, & \zeta^*_{2j-1} < t \leq \zeta^*_{2j}, \quad j \in \mathbb{Z}, \\
	\end{cases} 
	\end{equation}	
	is a relay function. In (\ref{relayfunc}), the sequence $ \{\zeta^*_j \} $ of switching moments is generated through the equation $ \zeta^*_{j}= j+\eta^*_{j}, \; j \in \mathbb{Z} $, where $ \{\eta^*_j \} $ is an unpredictable trajectory of the logistic map (\ref{logistic}).

One can confirm that $ \phi(t) $ is bounded such that $\displaystyle \sup_{t\in \mathbb R} |\phi(t)|\leq 3/4$. It was shown in paper [\citen{AkhmetPoincare}] that the function $ \phi(t) $ is the unique uniformly exponentially stable unpredictable solution of the differential equation
	\begin{equation} \label{UnpFunODEq}
	\vartheta' (t)= -2\vartheta(t)+\gamma^*(t).
	\end{equation}

It is not an easy task to visualize the unpredictable function $ \phi(t) $. Therefore, in order to illustrate the chaotic dynamics, we take into account the differential equation
	\begin{equation} \label{UnpFunODEq2}
	\vartheta' (t)= -2\vartheta(t)+\gamma(t),
	\end{equation}
	where
	\begin{equation}\label{relayfunc2}
	\gamma(t) =
	\begin{cases}
	1.5, & \zeta_{2j} < t \leq \zeta_{2j+1}, \quad j \in \mathbb{Z}, \\
	0.3, & \zeta_{2j-1} < t \leq \zeta_{2j}, \quad j \in \mathbb{Z}, \\
	\end{cases} 
	\end{equation}
and the sequence $ \{\zeta_j \} $ satisfies the equation $ \zeta_{j}= j+\eta_{j}, \; j \in \mathbb{Z} $, in which $ \{\eta_j \} $ is a solution of (\ref{logistic}) with $ \eta_0=0.4 $. The coefficient $ 3.91 $ used in the logistic map (\ref{logistic}) and the initial data $\eta_0=0.4$ were considered for shadowing analysis in the paper [\citen{Hammel}].

\begin{figure}[H]
		\centering
		\includegraphics[width=0.9\linewidth]{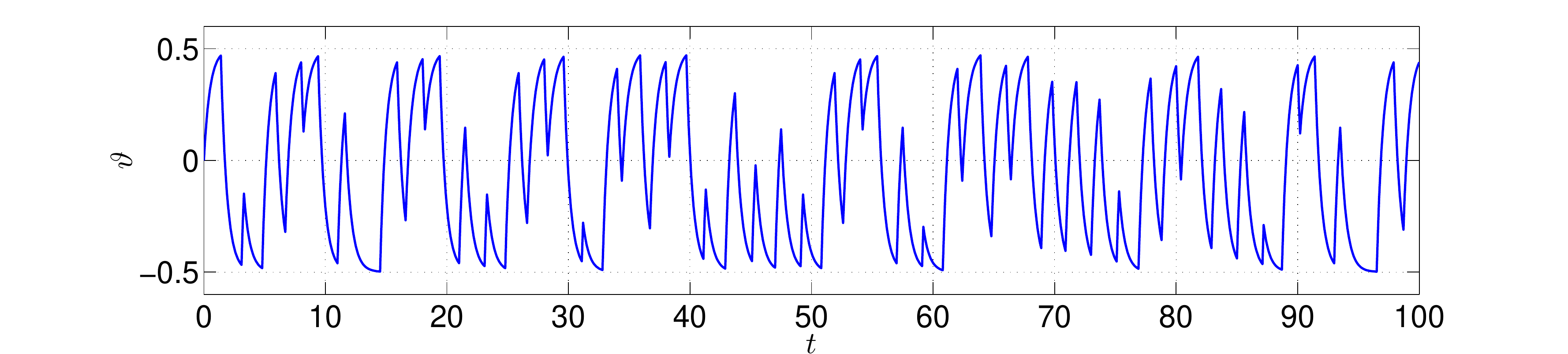}
		\caption{The solution of Equation (\ref{UnpFunODEq2}) with $ \vartheta(0)=0.3 $. The figure support that the function $ \phi(t) $ is unpredictable.}
		\label{UnpFunc}	
	\end{figure}

We depict in Figure \ref{UnpFunc} the solution of Equation (\ref{UnpFunODEq2}) with $ \vartheta(0)=0.3 $.  It is seen in Figure \ref{UnpFunc} that the behavior of the solution is irregular, and this support that the function $ \phi(t) $ is unpredictable.
	
	\subsection{Unpredictability Due to the Forcing Source Term}
		
	Let us perturb Equation (\ref{AdvectionEq}) with the unpredictable function $ \phi(t) $  defined by (\ref{UnpFunInEq}) and set up the equation
	
	\begin{equation} \label{PrturMainEq}
	\frac{\partial T}{\partial t} + u \frac{\partial T}{\partial x} + v \frac{\partial T}{\partial y} + w \frac{\partial T}{\partial z}= f(t,x,y,z,T)+\psi(\phi(t)),
	\end{equation}
	where $ u, v, w $, and $ f $ are in the form of (\ref{Velo&TfuncEq}) and $ \psi: [-3/4,3/4] \to \mathbb R $ is a continuous function.
	
	 Using the method of characteristics, one can reduce Equation (\ref{PrturMainEq})  to   system  (\ref{ParamSysEq}) that can be expressed in the form of (\ref{SysODEThmEq}) with
	\begin{equation*} \label{Eq2}
	G(X(t))=
	\begin{bmatrix}
	U \\
	V \\
	W \\
	F 
	\end{bmatrix}, \qquad 
	H(t)=
	\begin{bmatrix}
	0 \\
	0 \\
	0 \\
	\psi(\phi(t))
	\end{bmatrix}. 
	\end{equation*}
	According to the result of Theorem 5.2 [\citen{AkhmetPoincare}], if there exist positive constants $ L_1 $ and $ L_2 $ such that 
	\begin{eqnarray} \label{bilipschitz}
	L_1\left| s_1-s_2\right|  \leq \left|\psi(s_1)-\psi(s_2) \right| \leq L_2\left| s_1-s_2\right|
	\end{eqnarray} 
	for all $ s_1,s_2 \in [-3/4,3/4]$, then the function $H(t)$ is also unpredictable.

Now, in Equation (\ref{PrturMainEq}), let us take $u= -0.03 x+0.1 \sin(\frac{x}{80})+0.4,$ $v= -0.01 y-0.05 \sin(y),$ $w= -0.02  z+(0.05 \cos(y) - 0.00125 \cos(\frac{x}{80})) z$, $\psi(s)=6s$, and $f(t,x,y,z,T)= -1.5 T+0.1 w_2 T$. Since the conditions of Theorem \ref{Thm1} are valid and inequality (\ref{bilipschitz}) holds for these choices of $\psi$, $f$, $u$, $v$, and $w$, Equation (\ref{PrturMainEq}) exhibits Poincar\'{e} chaos. 

\begin{figure}[H]
\centering
\includegraphics[width=0.9\linewidth]{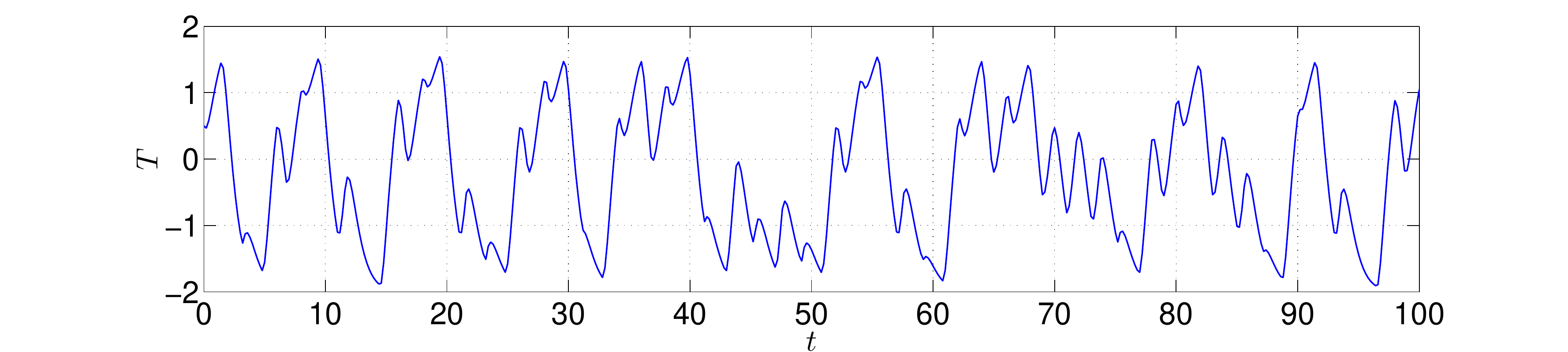}
\caption{The solution of Equation (\ref{PrturMainEq2}) with the initial condition $T(0, 0, 0, 0)= 0.5$. The figure reveals the presence of an unpredictable solution in the dynamics of (\ref{PrturMainEq}).}
\label{ChoBehSST}	
\end{figure}

In order to simulate the chaotic behavior, we consider the equation
\begin{equation} \label{PrturMainEq2}
\frac{\partial T}{\partial t} + u \frac{\partial T}{\partial x} + v \frac{\partial T}{\partial y} + w \frac{\partial T}{\partial z}= f(t,x,y,z,T)+\psi(\vartheta(t)),
\end{equation}
where $\vartheta(t)$ is the function depicted in Figure \ref{UnpFunc}, and $ u, v, w, f, \psi $ are the same as above. Figure \ref{ChoBehSST} shows  the solution $ T(t, x, y, z) $ of (\ref{PrturMainEq2}) corresponding to the initial condition $ T(0, 0, 0, 0)= 0.5 $. It is seen in Figure \ref{ChoBehSST} that the behavior of the solution is chaotic, and this supports the result of Theorem \ref{Thm1} such that Equation (\ref{PrturMainEq}) admits an unpredictable solution. 
 
Next, we will visualize the chaotic dynamics in the integral surface of SST. For that purpose, we omit the term of the meridional advection $ v \frac{\partial T}{\partial y} $ in (\ref{PrturMainEq}), which has less effect on SST compared with the zonal and vertical advections \cite{Bonjean}, and set up the equation
	\begin{equation} \label{2DimApproxAdv-PhiEq}
	\frac{\partial T}{\partial t} + u \frac{\partial T}{\partial x} + w \frac{\partial T}{\partial z}= -1.5 \, T+w \, T+6 \, \vartheta(t),
	\end{equation}
	where $u= 1.2+0.1 \sin(\frac{x}{80})+0.05 \sin(3t)$ and $w= 0.1 - 0.00125 \cos(\frac{x}{80}) z$. In (\ref{2DimApproxAdv-PhiEq}), $\vartheta(t)$ is again the function shown in Figure \ref{UnpFunc}.
	
\begin{figure}[H]
\centering
\includegraphics[width=0.6\linewidth]{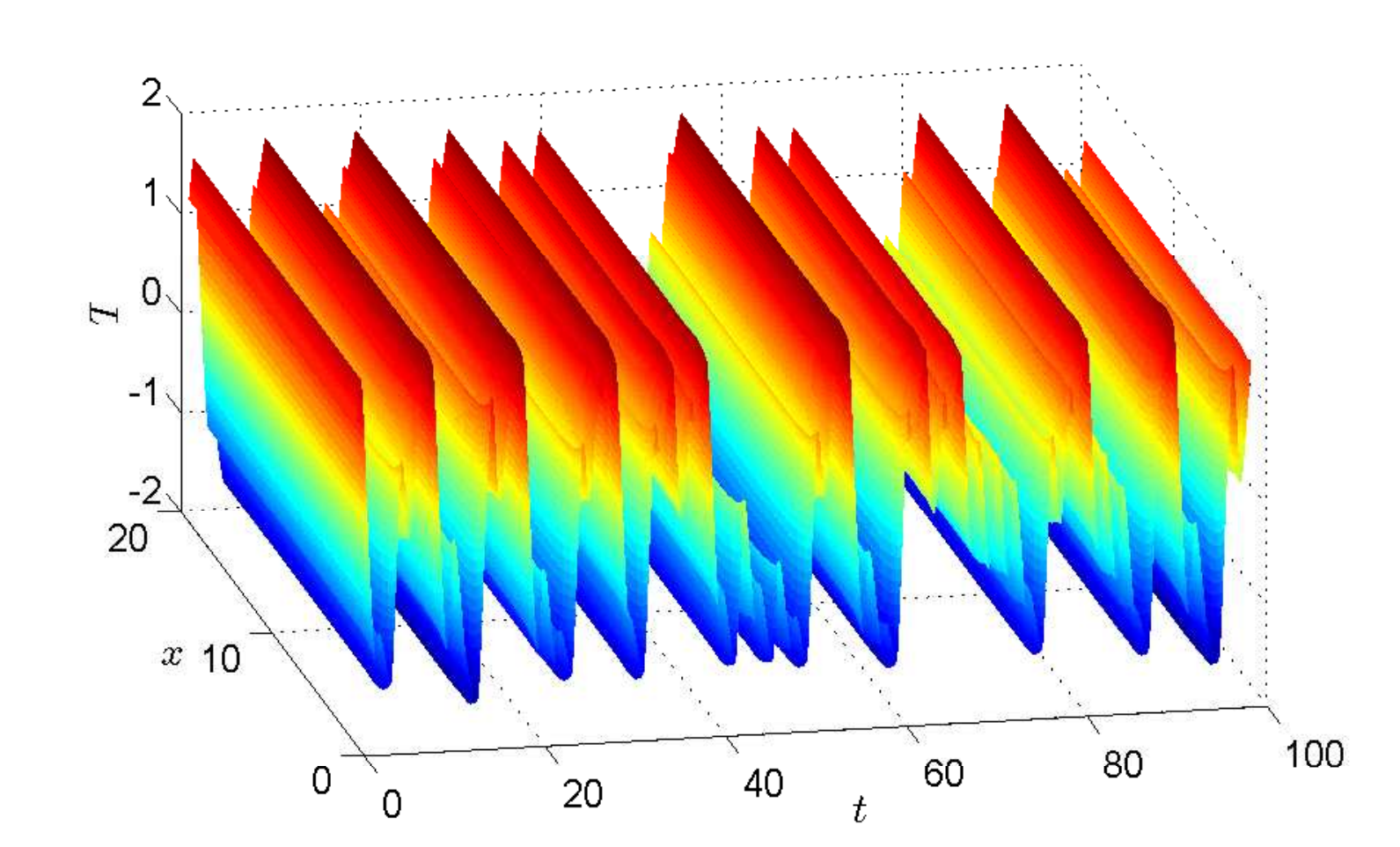}
\caption{The integral surface of (\ref{2DimApproxAdv-PhiEq}). The chaotic behavior in the SST is observable in the figure.}
\label{SSTSerfacePhi}	
\end{figure}

	We apply a finite difference scheme to solve Equation (\ref{2DimApproxAdv-PhiEq}) directly. In such a scheme, we need to specify boundary conditions along with an initial condition.	Using the initial condition $ T(0, x, z)= 5 $ and the boundary conditions $ T(t, 0, z)=T(t, x, 0)= 0.5$, we represent in Figure \ref{SSTSerfacePhi} the integral surface of (\ref{2DimApproxAdv-PhiEq}) with respect to $t,$ $x$, and the fixed value $z=0$ for $5\leq x\leq 20$ and $0\leq t \leq 100$. Figure \ref{SSTSerfacePhi} supports the result of Theorem \ref{Thm1} one more time such that Poincar\'{e} chaos is present in the dynamics. 

\subsection{Unpredictability Due to the Current Velocity}	
This subsection is devoted to the investigation of SST when the current velocity behaves chaotically. Here, we will make use of the unpredictable function $\phi(t)$ defined by (\ref{UnpFunInEq}) to apply perturbations to the zonal and vertical components of current velocity in Equation (\ref{AdvectionEq}).
	
We begin with considering the equation
\begin{equation} \label{E1ZonalPerturbEq}
\frac{\partial T}{\partial t} + [u +\psi(\phi(t))]\frac{\partial T}{\partial x} + v \frac{\partial T}{\partial y} + w \frac{\partial T}{\partial z}= f(t, x, y, z, T),
\end{equation}
where, in a similar way to (\ref{PrturMainEq}), $u, v, w$, and $f$ are in the form of (\ref{Velo&TfuncEq}), and $ \psi: [-3/4,3/4] \to \mathbb R $ is a continuous function.
	
One can confirm that Theorem \ref{Thm1} can be used to verify the existence of Poincar\'{e} chaos in the dynamics of (\ref{E1ZonalPerturbEq}) since it can be reduced by means of the method of characteristics to a system of the form (\ref{SysODEThmEq}) with	
	\begin{equation*}	
	H(t)=
	\begin{bmatrix}
		\psi(\phi(t)) \\
		0 \\
		0 \\
		0
	\end{bmatrix}, 
	\end{equation*}
	which is an unpredictable function provided that $\psi$ satisfies the condition (\ref{bilipschitz}). 

In order to demonstrate the chaotic dynamics of (\ref{E1ZonalPerturbEq}), we take $u= -0.003 \,x+0.2 \sin(\frac{{x}}{3})+0.4, $ $v= -0.001\,y$, $w= -0.002\,z- \frac{{0.2}}{3} \cos(\frac{{x}}{3})\,z$, $\psi(s)=3s$, $f=-1.5 \, T -3 \sin(3x)+0.2$, and consider the equation
	\begin{equation} \label{E1ZonalPerturbEq2}
	\frac{\partial T}{\partial t} + [u +\psi(\vartheta(t))]\frac{\partial T}{\partial x} + v \frac{\partial T}{\partial y} + w \frac{\partial T}{\partial z}= f(t, x, y, z, T),
	\end{equation}
where $\vartheta(t)$ is the function shown in Figure \ref{UnpFunc}.

\begin{figure}[H]
 		\centering
 		\includegraphics[width=0.9\linewidth]{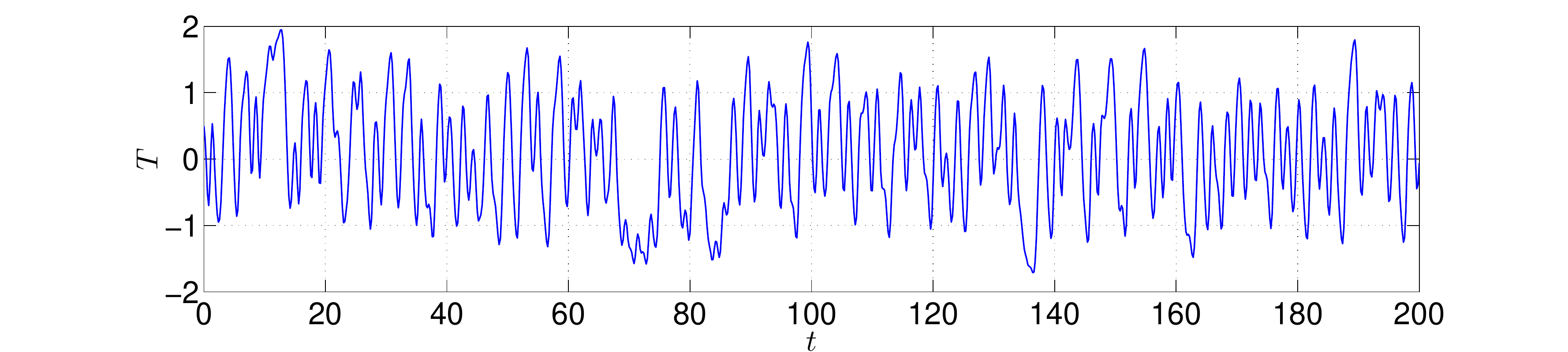}
 		\caption{The solution of (\ref{E1ZonalPerturbEq}) with $ T(0, 0, 0, 0)= 0.5 $. The chaotic behavior of the solution is apparent in the figure.}
 		\label{SSTWithUnpZonal-AlongCharac}	
 	\end{figure}

The time series of the solution of (\ref{E1ZonalPerturbEq2}) with $ T(0, 0, 0, 0)= 0.5 $ is depicted in Figure \ref{SSTWithUnpZonal-AlongCharac}. One can observe in the figure that the time series is chaotic, and this confirms the result of Theorem \ref{Thm1} such that Equation (\ref{E1ZonalPerturbEq}) possesses an unpredictable solution. More precisely, the perturbation of the zonal velocity component in Equation (\ref{AdvectionEq}) by the unpredictable function $ \psi(\phi(t)) $ affects the dynamics in such a way that the perturbed equation (\ref{E1ZonalPerturbEq}) is Poincar\'{e} chaotic.
		
Next, we will examine the case when the vertical velocity component in Equation (\ref{AdvectionEq}) is perturbed by the unpredictable function $\phi(t)$.
 For this aim we set up the equation	
\begin{equation} \label{E1meridionalPerturbEq3}
\frac{\partial T}{\partial t} + u\frac{\partial T}{\partial x} + v \frac{\partial T}{\partial y} + [w +\psi(\phi(t))] \frac{\partial T}{\partial z}= f(t, x, y, z, T),
\end{equation}
where the function $\psi: [-3/4,3/4] \to \mathbb R$ is continuous. If we take $ u= -0.001 \,x+0.2 \sin(\frac{{x}}{3})+0.4,$ $v=-0.001\,y,$ $w= -0.03 z- \frac{{0.2}}{3} \cos(\frac{{x}}{3}) z+3 \vartheta(t) $, $\psi(s)=3s$, and $ f= -1.7 \, T + 0.5 \, z+1.6 $, then Equation (\ref{E1meridionalPerturbEq3}) admits an unpredictable solution in accordance with Theorem \ref{Thm1}.

\begin{figure}[H]
\centering
\includegraphics[width=0.9\linewidth]{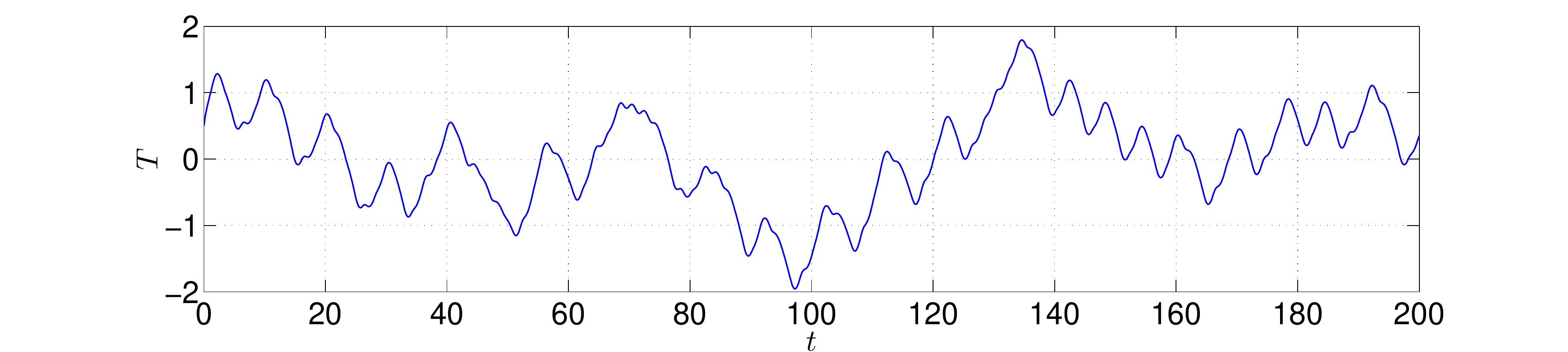}
\caption{Chaotic behavior of SST due to the perturbation of the vertical component of current velocity. The figure shows the solution of (\ref{E1meridionalPerturbEq4}) with $ T(0, 0, 0, 0)= 0.5 $.}
\label{SSTWithUnpVertical-AlongCharac}	
\end{figure}

We represent in Figure \ref{SSTWithUnpVertical-AlongCharac} the solution of the equation
\begin{equation} \label{E1meridionalPerturbEq4}
\frac{\partial T}{\partial t} + u\frac{\partial T}{\partial x} + v \frac{\partial T}{\partial y} + [w +\psi(\vartheta(t))] \frac{\partial T}{\partial z}= f(t, x, y, z, T),
\end{equation}
corresponding to the initial data $ T(0, 0, 0, 0)= 0.5 $. Here, we use the same $u,$ $v$, $w$, $\psi$, and $f$ as in (\ref{E1meridionalPerturbEq3}), and $\vartheta(t)$ is again the function whose time series is depicted in Figure \ref{UnpFunc}. The irregular fluctuations seen in the figure uphold the result of Theorem \ref{Thm1}.

\begin{figure}[H]
\subfigure[The integral surface of (\ref{E1ZonalPerturbEq}) at $ z= 0 $]{\includegraphics[width = 3.27in]{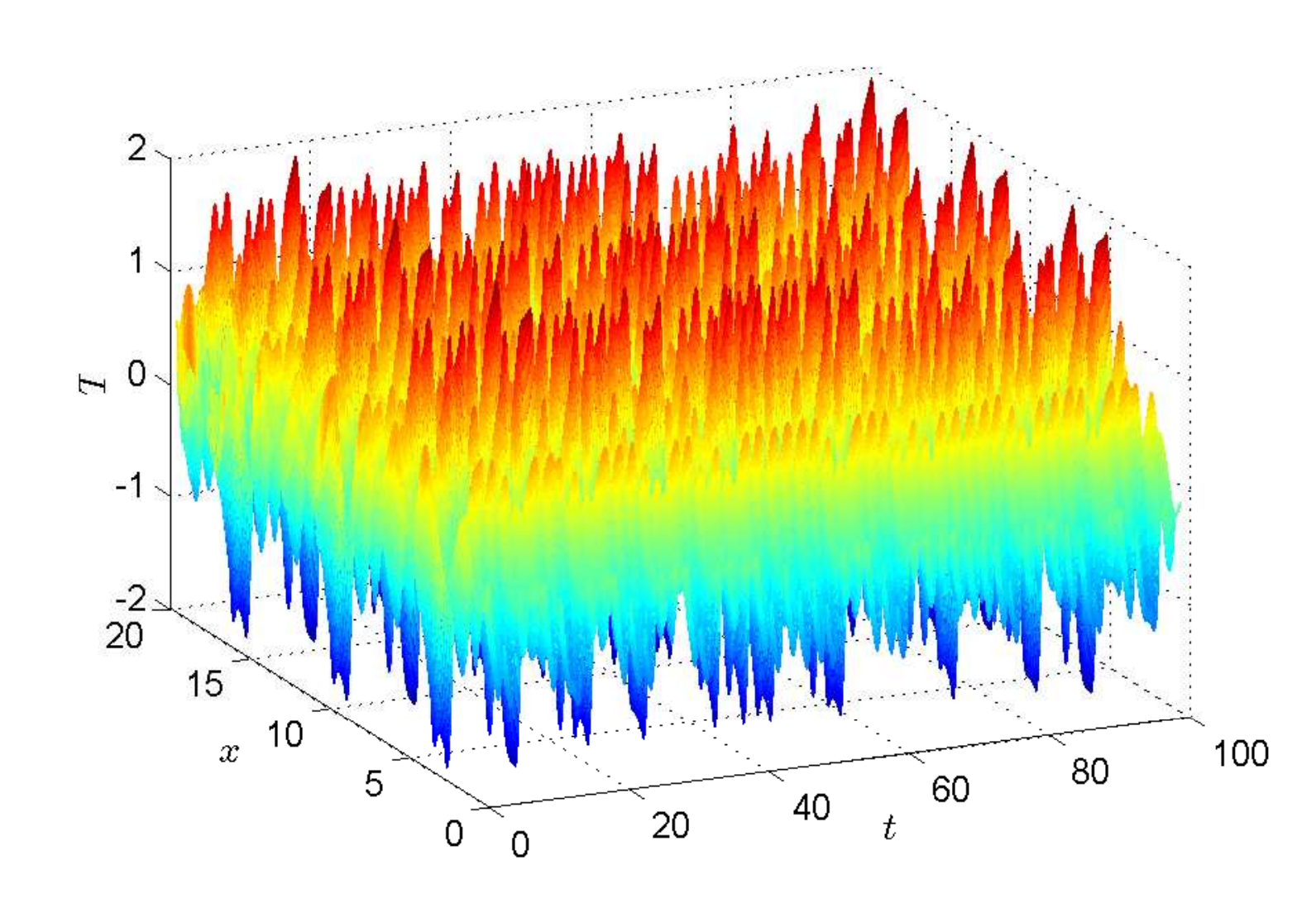}\label{SSTWithUnpZonal-Surf}}
\subfigure[The integral surface of (\ref{E1meridionalPerturbEq3}) at $ z= 1.5 $]{\includegraphics[width = 3.27in]{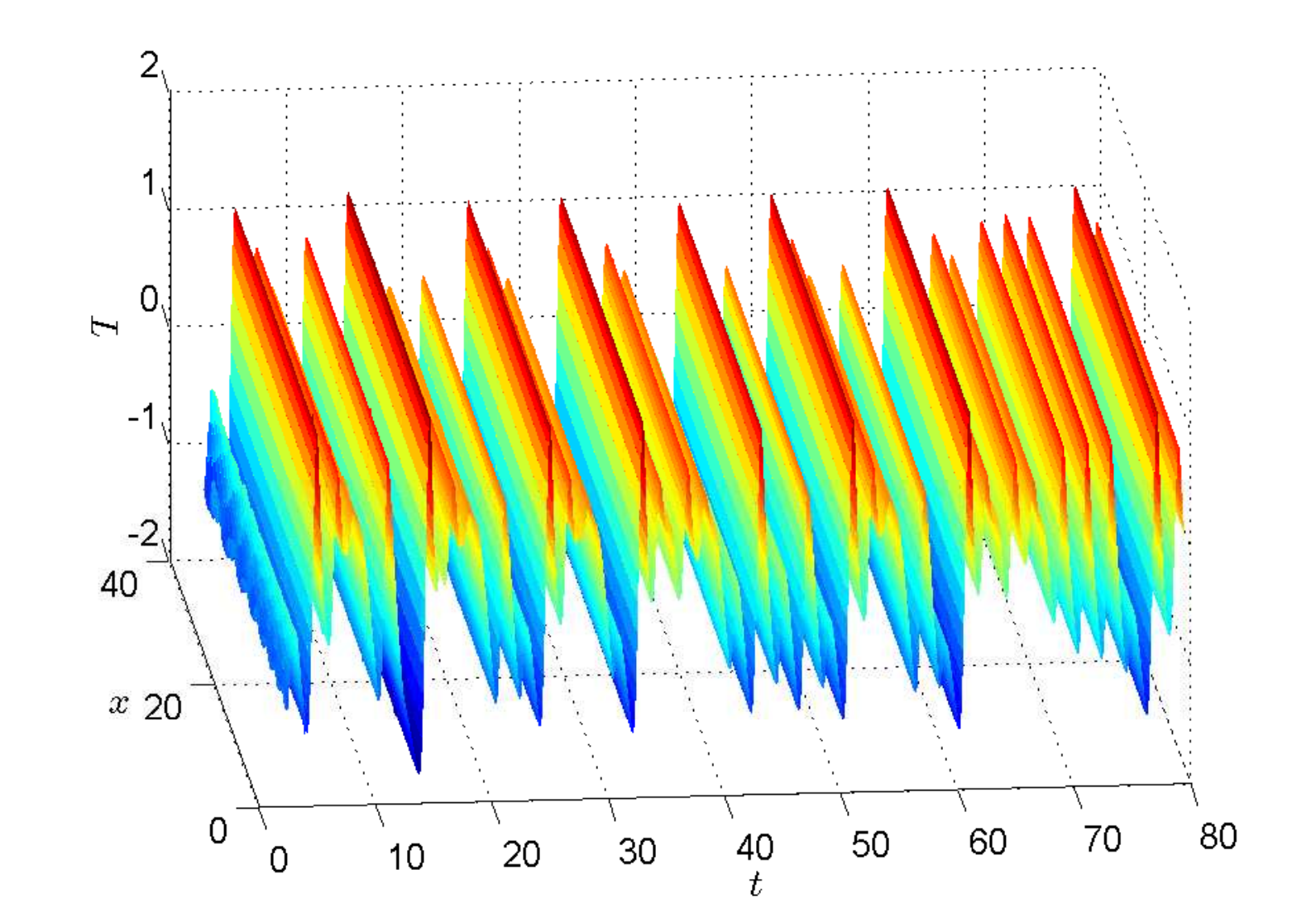}\label{SSTWithUnpVertical-Surf}}
\caption{Chaotic behavior of SST due to the current velocity with initial condition $ T(0, x, y, z)= \sin (xz) +1$, and boundary conditions $T(t, 0, y, z)=T(t, x, y, 0)= 0.5$. Both pictures in (a) and (b) reveal that chaotic behavior in the current velocity leads to the presence of chaos in SST.}
\label{CurrentVeloPerturb-SurfFg}   				
\end{figure}

We end up this subsection by illustrating	the influence of the chaotic current velocity on the integral surface of SST. Figure \ref{CurrentVeloPerturb-SurfFg} (a) shows the integral surface of (\ref{E1ZonalPerturbEq}) with 
$u= 1.5+0.5 \sin x, $ $v=0$, $w= 1-0.5 \cos x$, $\psi(s)=2s$, and $f=-1.2 \, T -3 \sin(3x)$ at $z=0$. The initial condition $ T(0, x, y, z)= \sin (xz) +1$ and the boundary conditions $ T(t, 0, y, z)=T(t, x, y, 0)= 0.5 $ are utilized in the simulation. One can see in Figure \ref{CurrentVeloPerturb-SurfFg} (a) that the SST has chaotic behavior in keeping with the result of Theorem \ref{Thm1}. On the other hand, using the same initial and boundary conditions, we represent in Figure \ref{CurrentVeloPerturb-SurfFg} (b) the integral surface of (\ref{E1meridionalPerturbEq3}) with $u= 1, $ $v=0$, $w= 1$, $\psi(s)=2s$, and $f=-1.2 \, T +3 \sin(3z)$ at $z=1.5$. Figure \ref{CurrentVeloPerturb-SurfFg} (b) also manifests that the applied perturbation on the vertical component of current velocity make the Equation (\ref{E1meridionalPerturbEq3}) behave chaotically even if it is initially non-chaotic in the absence of the perturbations.

\section{Extension of Chaos through the Globe Ocean } \label{HorizontalExtension}
	
Chaotic behavior may transmit from one model to another \cite{AkhmetReplication}. This transmission interprets, for instance, why the unpredictability in one stock market or in the weather of one area is affected by another. Chaos in SST may be gained from another endogenous chaotic system like air temperature or wind speed. We can deal with the global ocean as a finite union of subregions. Each of these subregions may be controlled by different models depending on the position and circumstances. An assumption of the existence of chaotic and non-chaotic subregions for SST behavior is very probable. However, it seems quite unreasonable to imagine a predictable SST for one region whereas its neighbor region is characterized by an unpredictable SST. The mutual effect in SST between neighbor regions can be seen by coupling their controlling models.  
	
\subsection{Coupling of Advection Equations}	
	
In this part of the paper we deal with the extension of chaos in coupled advection equations. For that purpose, we consider a Poincar\'{e} chaotic advection equation of the form (\ref{PrturMainEq}) as the drive, and we take into account the equation
\begin{equation} \label{response_advec_eq}
\frac{\partial \tilde{T}}{\partial t} + \tilde{u} \frac{\partial \tilde{T}}{\partial x} + \tilde{v} \frac{\partial \tilde{T}}{\partial y} + \tilde{w} \frac{\partial \tilde{T}}{\partial z}= \tilde{f}(t,x,y,z,\tilde{T}) + g(T)
\end{equation}
as the response, in which $g$ is a continuous function and $T$ is a solution of the drive equation (\ref{PrturMainEq}). 
We assume that the response does not possess chaos in the absence of the perturbation, i.e., we suppose that the advection equation
\begin{equation} \label{response_advec_eq2}
\frac{\partial \tilde{T}}{\partial t} + \tilde{u} \frac{\partial \tilde{T}}{\partial x} + \tilde{v} \frac{\partial \tilde{T}}{\partial y} + \tilde{w} \frac{\partial \tilde{T}}{\partial z}= \tilde{f}(t,x,y,z,\tilde{T})
\end{equation}	
is non-chaotic.	
	
%
%
%
%
	
\begin{figure}[H]
		\centering
		\includegraphics[width=0.9\linewidth]{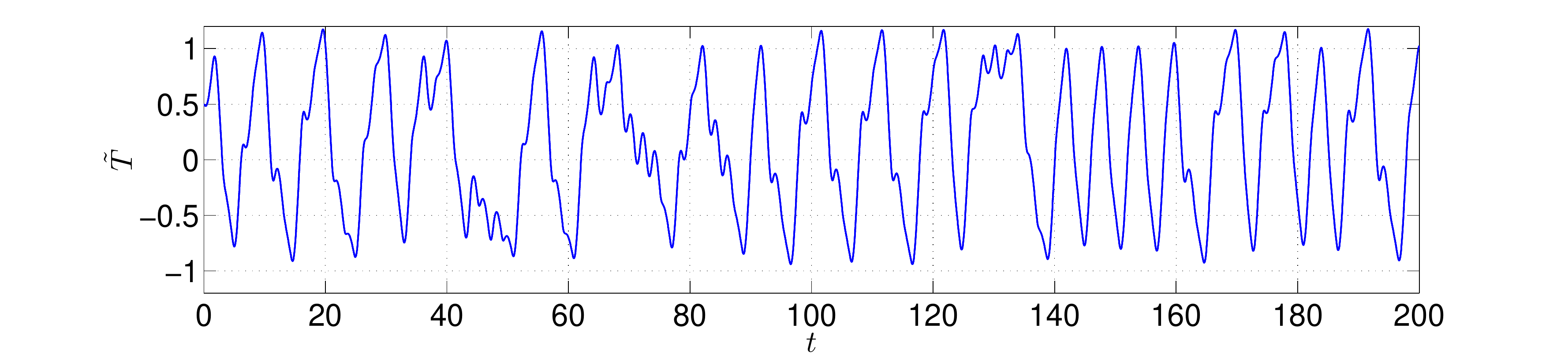}
		\caption{The solution of the response equation (\ref{response_advec_eq}) with initial condition $T(0, 0, 0, 0)= 0.5$. The figure manifests the extension of chaos in the coupled system (\ref{PrturMainEq})-(\ref{response_advec_eq}).}
		\label{ChaosExten}	
	\end{figure}

To demonstrate the extension of chaos numerically, let us consider the response equation (\ref{response_advec_eq}) with $u=1.2$, $v=0$, $w=0.3$, $f=-1.5 \tilde{T} + 0.2$, and $g(T)=T$. Using the solution $T$ of Equation (\ref{2DimApproxAdv-PhiEq}) satisfying $T(0, 0, 0, 0)= 0.5$ as the perturbation in Equation (\ref{response_advec_eq}), we depict in Figure \ref{ChaosExten} the solution $\tilde{T}$ of (\ref{response_advec_eq}) corresponding to the initial data $\tilde{T}(0, 0, 0, 0)= 0.5$. Figure \ref{ChaosExten} reveals the extension of chaos in the coupled system (\ref{PrturMainEq})-(\ref{response_advec_eq}).

\subsection{Coupling of the Advection Equation with Vallis Model}
	
The Lorenz-like form of the Vallis model is given by \cite{Vallis}
\begin{equation} \label{LorenzFormVallis}
\begin{split}
& \frac{du}{dt} =B\,T_d-C\,u,\\
& \frac{dT_d}{dt} = u\,T_s-T_d ,  \\	
& \frac{dT_s}{dt} = -u\,T_d-T_s+1,
\end{split}
\end{equation}
where $ u $ represents the zonal velocity, $ T_d=(T_e-T_w)/2 $, $ T_s=(T_e+T_w)/2 $, $ T_e $ and $ T_w $ are the SST in the eastern and western ocean respectively, and $B$ and $C$ are constants. It was shown in paper [\citen{Vallis}] that system (\ref{LorenzFormVallis}) with the parameters $B = 102 $ and $C = 3$ is chaotic. The existence of chaos in the dynamics of Vallis systems was also investigated in the studies [\citen{Garay,Borghezan}].

Next, we take into account the equations	
	\begin{equation} \label{CoupledAdvecLorenz-Td.Eq1}
	\frac{\partial T_1}{\partial t} + 1.2 \frac{\partial T_1}{\partial x} + 0.3 \frac{\partial T_1}{\partial z}= -1.2\, T_1 - 1 + 2 \sin x,
	\end{equation}
	\begin{equation} \label{CoupledAdvecLorenz-u.Eq1}
	\frac{\partial T_2}{\partial t} + 1.2 \frac{\partial T_2}{\partial x} + 0.3 \frac{\partial T_2}{\partial z}= -2 \, T_2 + 4 \sin x,
	\end{equation}
	\begin{equation} \label{CoupledAdvecLorenz-Td-u.Eq1}
	\frac{\partial T_3}{\partial t} + 0.6 \frac{\partial T_3}{\partial x} + 0.5 \frac{\partial T_3}{\partial z}= -2\, T_3 -1 + 3 \sin x,
	\end{equation}
and
	\begin{equation} \label{2CoupledAdvecLorenz.Eq1}
	\frac{\partial T_4}{\partial t} + 1.2 \frac{\partial T_4}{\partial x} + 0.3 \frac{\partial T_4}{\partial z}= -1.5 \, T_4.
	\end{equation}	
One can verify that the equations (\ref{CoupledAdvecLorenz-Td.Eq1}), (\ref{CoupledAdvecLorenz-u.Eq1}), (\ref{CoupledAdvecLorenz-Td-u.Eq1}), and (\ref{2CoupledAdvecLorenz.Eq1}) are all non-chaotic such that they admit asymptotically stable regular solutions. By applying perturbations to these equations, we set up the following ones:
	\begin{equation} \label{CoupledAdvecLorenz-Td.Eq}
	\frac{\partial T_1}{\partial t} + 1.2 \frac{\partial T_1}{\partial x} + 0.3 \frac{\partial T_1}{\partial z}= -1.2\, T_1 - 1 + 2 \sin x +4.6 \, T_s,
	\end{equation}
	\begin{equation} \label{CoupledAdvecLorenz-u.Eq}
	\frac{\partial T_2}{\partial t} + (1.2+0.8 \, u) \frac{\partial T_2}{\partial x} + 0.3 \frac{\partial T_2}{\partial z}= -2 \, T_2 + 4 \sin x,
	\end{equation}
	\begin{equation} \label{CoupledAdvecLorenz-Td-u.Eq}
	\frac{\partial T_3}{\partial t} + (0.6+ u) \frac{\partial T_3}{\partial x} + 0.5 \frac{\partial T_3}{\partial z}= -2\, T_3 -1 + 3 \sin x +4 \, T_s,
	\end{equation}
	\begin{equation} \label{2CoupledAdvecLorenz.Eq}
	\frac{\partial T_4}{\partial t} + 1.2 \frac{\partial T_4}{\partial x} + 0.3 \frac{\partial T_4}{\partial z}= -1.5 \, T_4 + 2.7 \, T_2,
	\end{equation}
where $(u, T_d, T_s)$ is the solution of the chaotic Vallis model (\ref{LorenzFormVallis}) with $B = 102 $, $C = 3$ and the initial conditions $u(0)=2,$ $T_d(0)=0.2$, and $T_s(0)=0.4$.

In Equation (\ref{CoupledAdvecLorenz-Td.Eq}) the forcing term is perturbed by the SST average, $T_s$, whereas in Equation (\ref{CoupledAdvecLorenz-u.Eq}) the zonal velocity of Vallis model, $u$, is used as perturbation. On the other hand, in Equation (\ref{CoupledAdvecLorenz-Td-u.Eq}) both the forcing term and the zonal velocity components are perturbed with the solution of (\ref{LorenzFormVallis}). Moreover, the solution $T_2$ of (\ref{CoupledAdvecLorenz-u.Eq}) is used as a perturbation in the forcing term of Equation (\ref{2CoupledAdvecLorenz.Eq}). 

%
%
%
%

\begin{figure}[H]
\subfigure[]{\includegraphics[width = 6.4in]{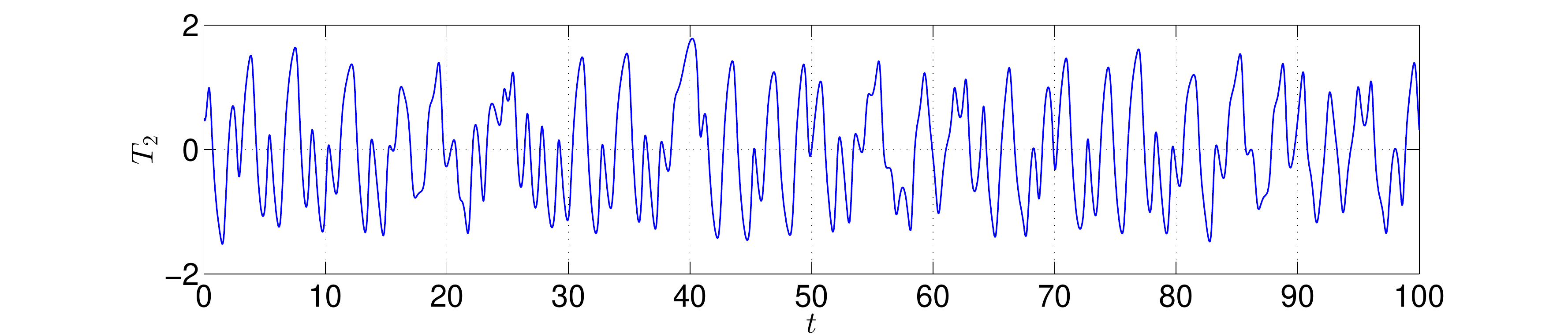}\label{SSTWithVallispZonal-AlongCharac}}
\subfigure[]{\includegraphics[width = 6.4in]{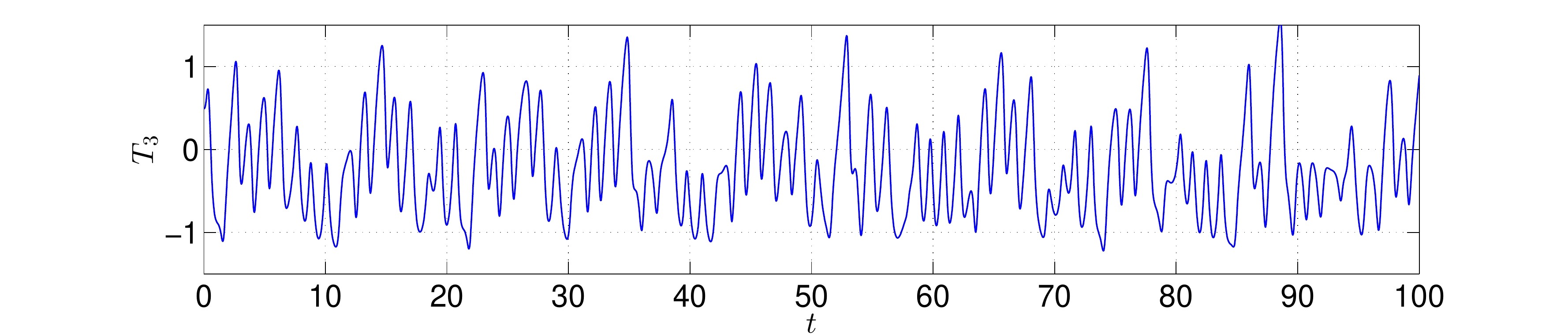}\label{SSTWithVallispZonal-T-AlongCharac}}
\caption{The extension of the chaotic behavior by Equations (\ref{CoupledAdvecLorenz-u.Eq}) and (\ref{CoupledAdvecLorenz-Td-u.Eq}). (a) The time series of the solution of Equation (\ref{CoupledAdvecLorenz-u.Eq}), (b) The time series of the solution of Equation (\ref{CoupledAdvecLorenz-Td-u.Eq}). The initial data $T_2(0, 0, 0, 0)= 0.5$ and $T_3(0, 0, 0, 0)= 0.5$ are used.}  
\label{CoupledAdvecLorenzAlongCharacFg}   	
\end{figure}

Figure \ref{CoupledAdvecLorenzAlongCharacFg} (a) and (b) respectively show the solutions $T_2$, $T_3$ of Equations (\ref{CoupledAdvecLorenz-u.Eq}) and (\ref{CoupledAdvecLorenz-Td-u.Eq}), respectively. The initial data $ T_2(0, 0, 0, 0)= 0.5 $ and $ T_3(0, 0, 0, 0)= 0.5 $ are used in the simulation. Figure \ref{CoupledAdvecLorenzAlongCharacFg} reveals that the chaos of the model (\ref{LorenzFormVallis}) is extended by  Equations (\ref{CoupledAdvecLorenz-u.Eq}) and (\ref{CoupledAdvecLorenz-Td-u.Eq}).

\begin{figure}[H]
     	\subfigure[]{\includegraphics[width = 3.27in]{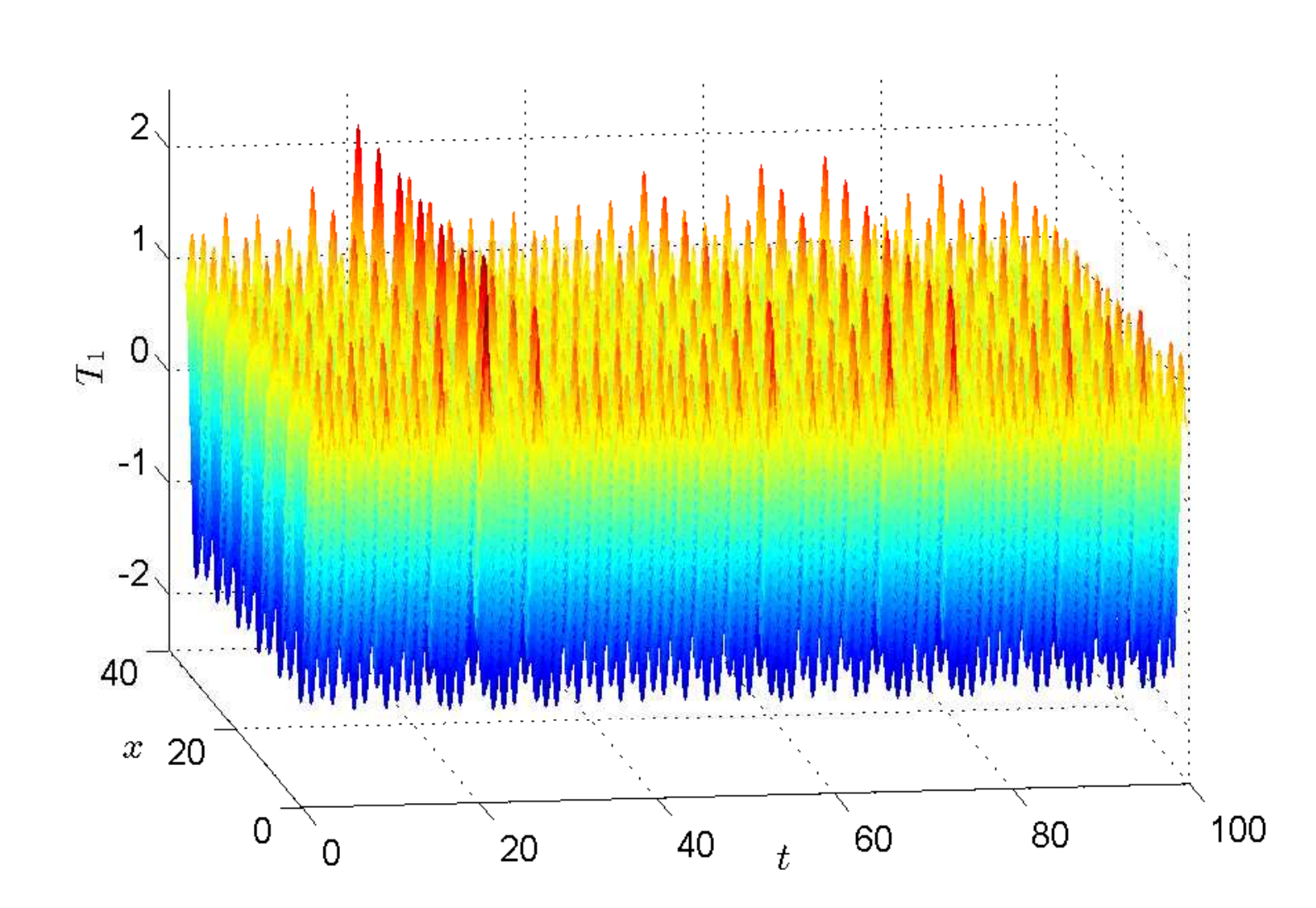}\label{SSTWithVallisTs-Surf}}
     	\subfigure[]{\includegraphics[width = 3.27in]{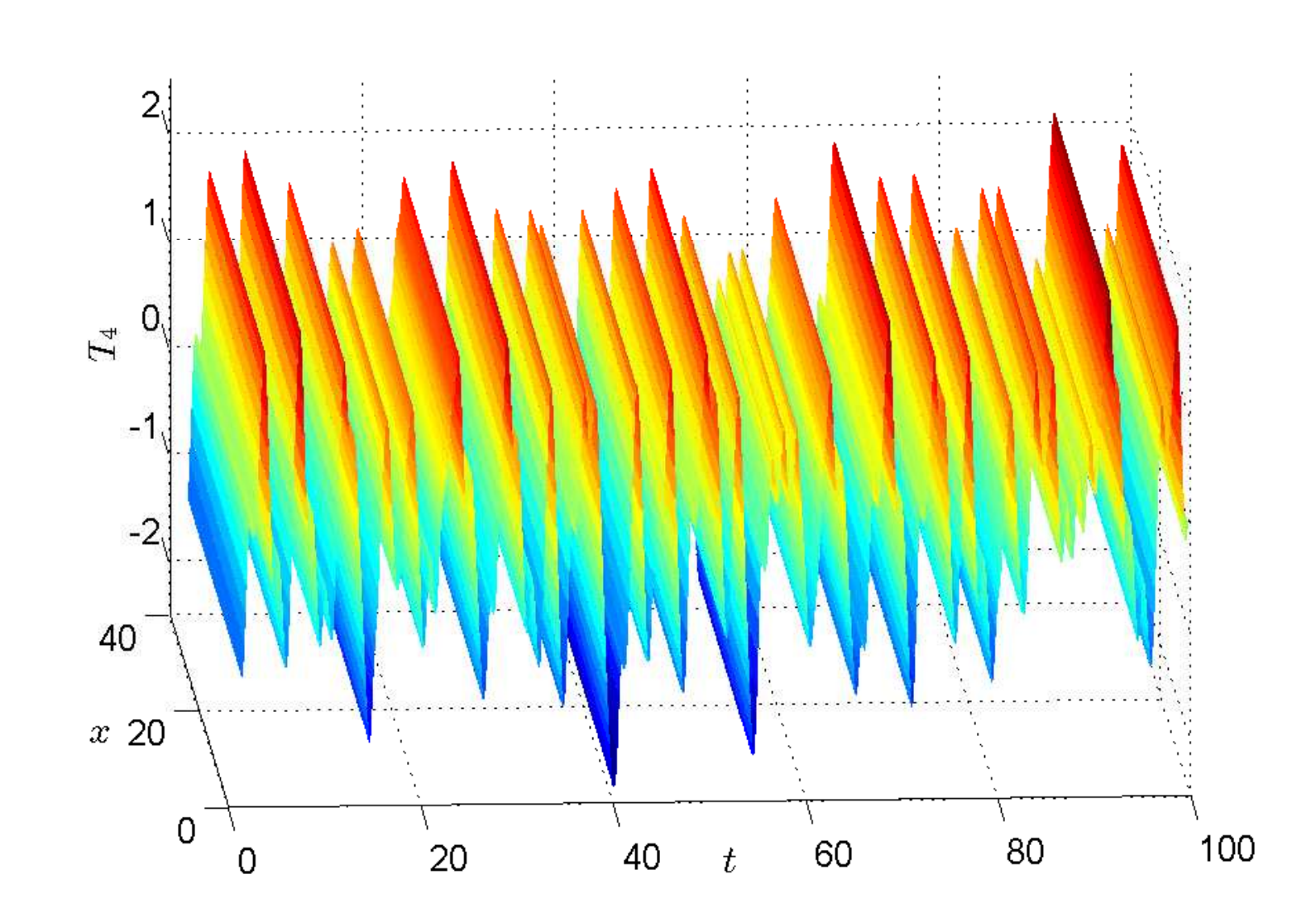}\label{SSTWithVallispZonal-Surf}}
     	\caption{Extension of chaos by Equations (\ref{CoupledAdvecLorenz-Td.Eq}) and (\ref{2CoupledAdvecLorenz.Eq}). (a) The integral surface of Equation (\ref{CoupledAdvecLorenz-Td.Eq}), (b) The integral surface of Equation (\ref{2CoupledAdvecLorenz.Eq}).}
     	\label{CoupledAdvecLorenzSurfFg}   	
     \end{figure}

On the other hand, we depict in Figure \ref{CoupledAdvecLorenzSurfFg} (a) and (b) the $3$ dimensional integral surfaces corresponding to Equations (\ref{CoupledAdvecLorenz-Td.Eq}) and  (\ref{2CoupledAdvecLorenz.Eq}), respectively. Here, we make use of the boundary conditions $T_1(0, x, z)=T_1(t, 0, z)=T_1(t, x, 0)= 0.5$ and $T_4(0, x, z)=T_4(t, 0, z)=T_4(t, x, 0)= 0.5$. The figure confirms one more time that the chaos of system (\ref{LorenzFormVallis}) is extended.     
    
\subsection{Coupling of Vallis Models}

Our purpose in this subsection is to demonstrate numerically our suggestion that chaos can be extended between the regions of some global climate variabilities. We assume that there are intermediate subregions located between these main regions and chaos can transmit from one region to another in a sequential way.
    
We also suggest that the IOD can be described by a Vallis model in the form of (\ref{LorenzFormVallis}) with parameters appropriate to the Indian Ocean. Evaluation of these parameters is rather difficult. However, for simplicity we can choose these values such that system (\ref{LorenzFormVallis}) does not exhibit chaotic behavior. Similar arguments can also be supposed for the AMO and SAM. 

%
%
%
   
\begin{figure}[H]
    	\subfigure[]{\includegraphics[width = 6.4in]{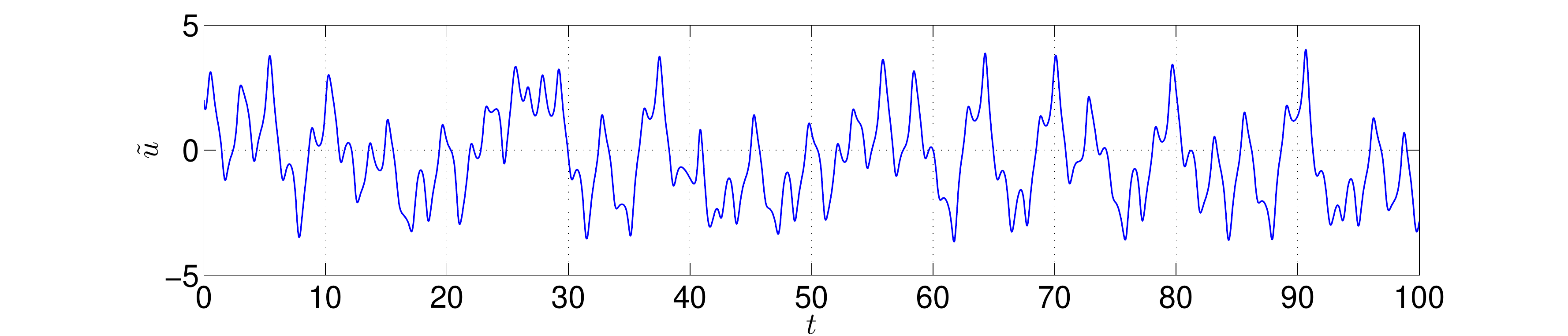}\label{ENSO-IOD-u}} \\
    	\subfigure[]{\includegraphics[width = 6.4in]{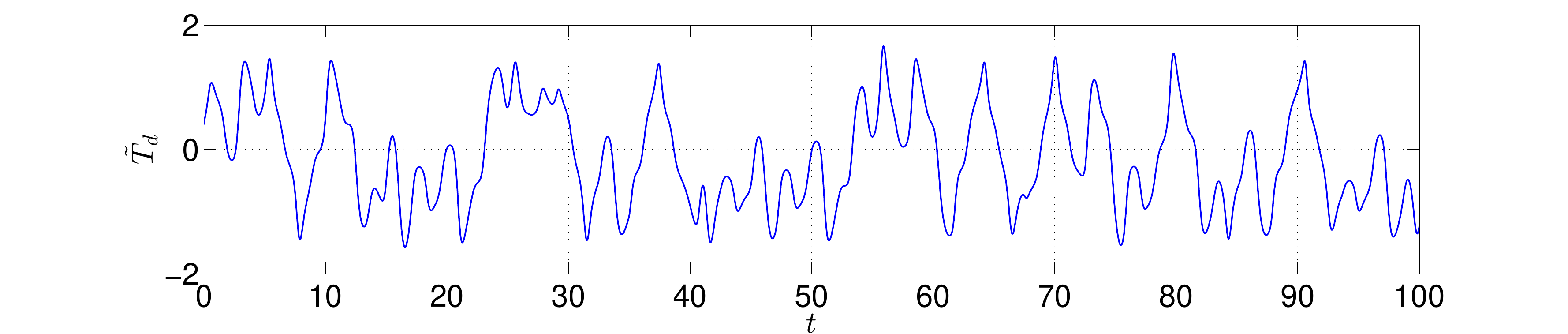}\label{ENSO-IOD-Td}}
    	\subfigure[]{\includegraphics[width = 6.4in]{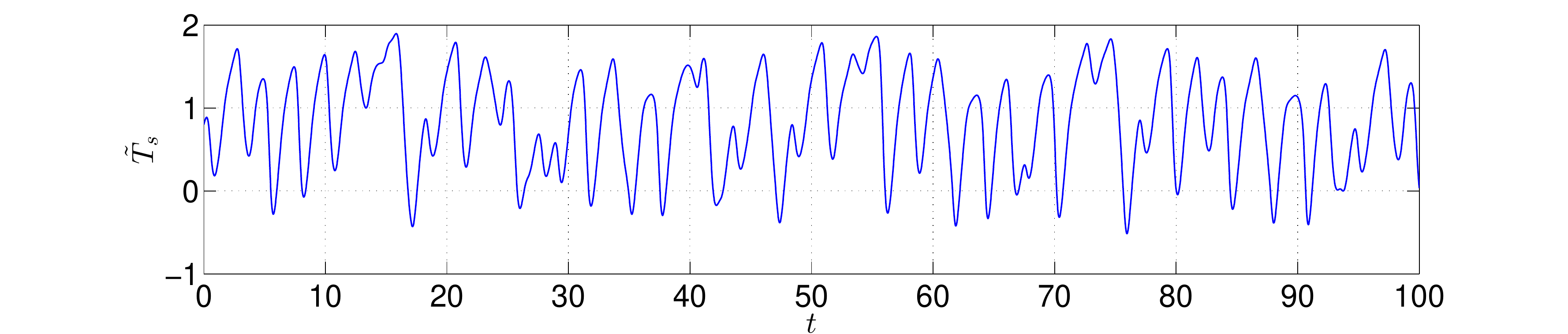}\label{ENSO-IOD-Ts}}
    	\caption{The solution of system (\ref{LorenzVallisIOD}) which reveals chaos extension between a pair of Vallis systems}
    	\label{CoupledTwoVallis}   	
    \end{figure}

To demonstrate the extension of chaos, let us consider the perturbed Vallis system
\begin{equation} \label{LorenzVallisIOD}
\begin{split}
& \frac{d \tilde u}{dt} = \tilde B\, \tilde T_d- \tilde C\, \tilde u +1.5 \, u,\\
& \frac{d \tilde T_d}{dt} = \tilde u\, \tilde T_s- \tilde T_d +0.3 \, T_d,  \\	
& \frac{d \tilde T_s}{dt} = - \tilde u\, \tilde T_d- \tilde T_s+1 +0.2 \, T_s,
\end{split}
\end{equation}
where $(u, T_d, T_s)$ is the solution of the chaotic Vallis system (\ref{LorenzFormVallis}) with $B = 102$ and $C = 3$ corresponding to the initial conditions $u(0)=2,$ $T_d(0)=0.2$ and $T_s(0)=0.4$.
We use the parameters $\tilde B = 20$ and $\tilde C = 7$ in (\ref{LorenzVallisIOD}) and assume that the unperturbed Vallis model 
\begin{equation} \label{VallisIOD2}
\begin{split}
& \frac{d \tilde u}{dt} = \tilde B\, \tilde T_d- \tilde C\, \tilde u, \\
& \frac{d \tilde T_d}{dt} = \tilde u\, \tilde T_s- \tilde T_d,  \\	
& \frac{d \tilde T_s}{dt} = - \tilde u\, \tilde T_d- \tilde T_s+1
\end{split}
\end{equation}
represents the IOD with these parameter values. Utilizing the initial conditions $\tilde u(0)=2$, $\tilde T_d(0)=0.2$, and $ \tilde T_s(0)=0.4$, we represent in Figure \ref{CoupledTwoVallis} the time series of $\tilde u,$ $\tilde T_d$, and $\tilde T_s$ coordinates of the solution of system (\ref{LorenzVallisIOD}). One can see in Figure \ref{CoupledTwoVallis} that system (\ref{LorenzVallisIOD}) possesses chaotic behavior.
 
	\section{Ocean-Atmosphere Unpredictability Interaction} \label{VerticalExtension}
	
    In this section, we discuss the possibility of the ``vertical'' extension of unpredictability, i.e. the transmission of chaotic dynamics from ocean to atmosphere and vice versa. To demonstrate this interaction we apply the Lorenz system (\ref{LorenzSystem}) for the atmosphere and the Vallis model (\ref{LorenzFormVallis}) for the ocean. Vallis model is constructed for the domain length of $ 7500 \, \text{km} $, however, depending on the method of construction, the model can be applied for more localized region to be compatible with the Lorenz model. 
    
%
%
%
%
%
%
%
%
%
      
Heat and momentum exchanges are two important ways of interaction between ocean and atmosphere. The heat exchange is mainly controlled by the air-sea temperature gradient, and, on the other hand, the momentum transfer is determined by the sea-surface stress caused by wind and currents \cite{Gallego}. These characteristics are represented in both  Lorenz system (\ref{LorenzSystem}) and Vallis model (\ref{LorenzFormVallis}). Two coordinates in the Lorenz system represent temperature, whereas the third one is related to velocity, and the same could be said for the Vallis system. Therefore, the interaction between ocean and atmosphere can be modeled by coupling the Lorenz and Vallis models.

Let us consider the coupled Lorenz-Vallis systems    
    \begin{equation} \label{GenLorenzVallisCouple}
    \begin{split}
    & \frac{dx}{dt} =\sigma(y-x) + f_1(u,T_d,T_s),\\
    & \frac{dy}{dt} = x(r-z)-y+f_2(u,T_d,T_s),  \\	
    & \frac{dz}{dt} = xy - b \,z+f_3(u,T_d,T_s),
    \end{split}
    \end{equation}
and    
    \begin{equation} \label{GenVallisLorenzCouple}
    \begin{split}
    & \frac{du}{dt} =B\,T_d-C\,u+g_1(x,y,z),\\
    & \frac{dT_d}{dt} = u\,T_s-T_d+g_2(x,y,z),  \\	
    & \frac{dT_s}{dt} = -u\,T_d-T_s+1+g_3(x,y,z),
    \end{split}
    \end{equation}
where $f_i$, $g_i$, $i=1,2,3$, are continuous functions. The coupled model (\ref{GenLorenzVallisCouple})--(\ref{GenVallisLorenzCouple}) is in a sufficiently general form of interaction between the atmosphere and the ocean, where the functions $ f_i, g_i, \, i=1, 2, 3 $ are given in most general form. 

To demonstrate the transmission of chaos between the atmosphere and ocean, we consider specific forms of the coupled model (\ref{GenLorenzVallisCouple})--(\ref{GenVallisLorenzCouple}). This technique relies on the theoretical investigations of replication of chaos introduced in [\citen{AkhmetReplication}]. 

In the case of upward transmission of chaos from the ocean to the atmosphere, we consider (\ref{GenLorenzVallisCouple}) with specific choices of the perturbation functions $ f_1, f_2 $ and $ f_3 $ to set up the following system,
\begin{equation} \label{LorenzVallisCouple}
\begin{split}
& \frac{dx}{dt} =\sigma(y-x) + 3 \sin u,\\
& \frac{dy}{dt} = x(r-z)-y+6\,T_d,  \\	
& \frac{dz}{dt} = xy - b \,z+0.5\,T_s^2,
\end{split}
\end{equation}
where $(u, T_d, T_s)$ is the solution of the chaotic Vallis system (\ref{LorenzFormVallis}) with $ B = 102 $, $ C = 3 $ and the initial data $u(0)=2,$ $T_d(0)=0.2$, $T_s(0)=0.4$. We use the parameter values $ \sigma = 10 $, $ r = 0.35 $ and $ b = 8/3 $ in (\ref{LorenzVallisCouple}) such that the corresponding unperturbed Lorenz system (\ref{LorenzSystem}) does not possess chaos \cite{Sparrow}. 

\begin{figure}[H]
    	\subfigure[]{\includegraphics[width = 6.4in]{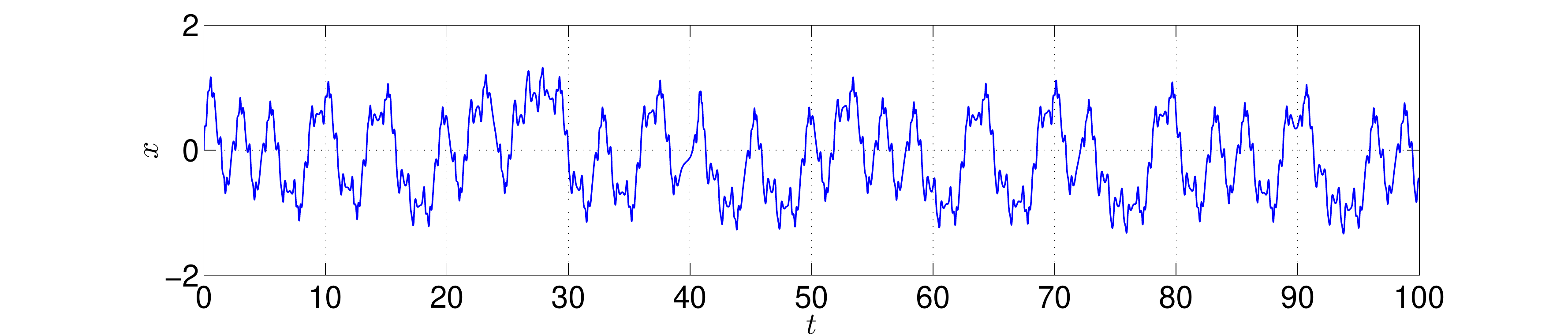}\label{LorVall-x}} \\
    	\subfigure[]{\includegraphics[width = 6.4in]{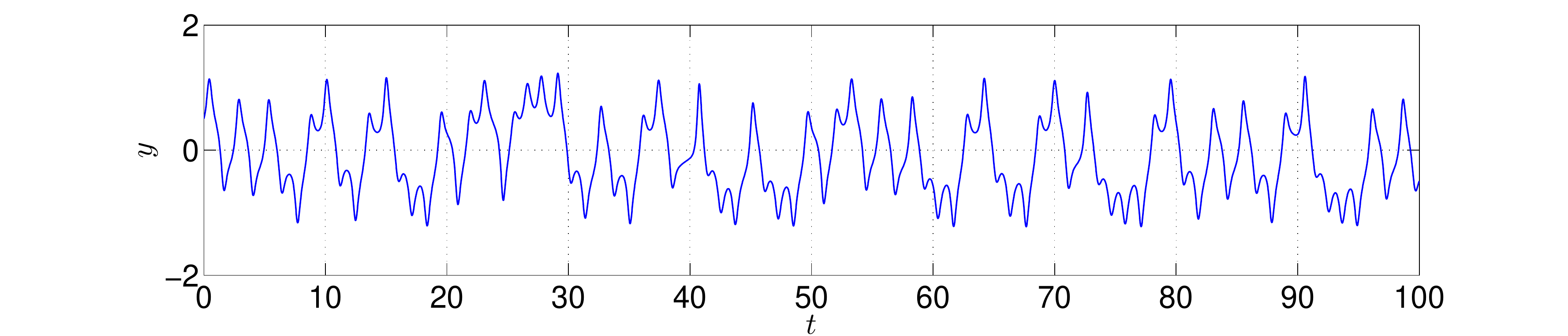}\label{LorVall-y}}
    	\subfigure[]{\includegraphics[width = 6.4in]{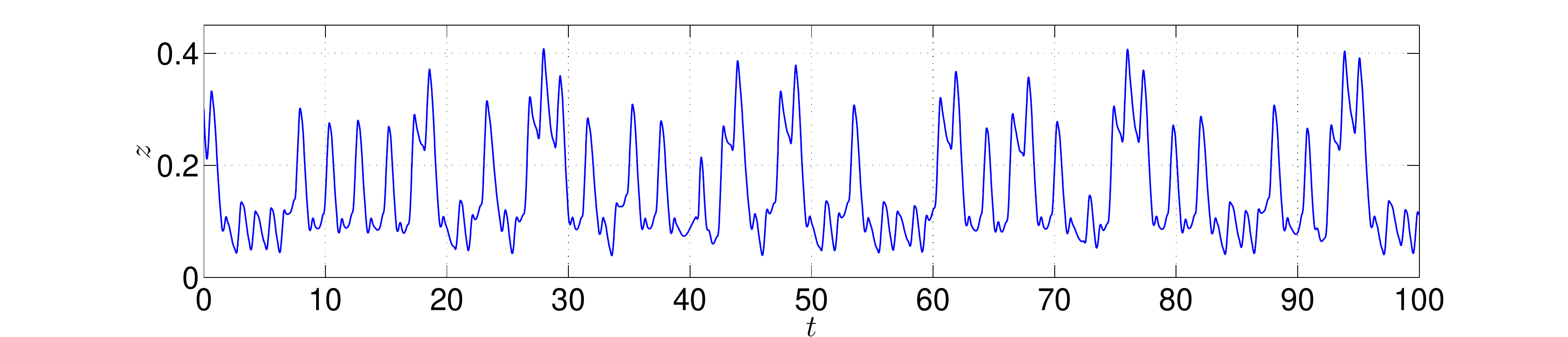}\label{LorVall-z}}
    	\caption{The chaotic solution of the perturbed Lorenz system (\ref{LorenzVallisCouple})}
    	\label{CoupledLorenzVallis}   	
    \end{figure}

Figure \ref{CoupledLorenzVallis} shows the time series of the $x$, $y$, and $z$ components of the solution of system (\ref{LorenzVallisCouple}). The initial data $ x(0)=0,$ $y(0)=0.5$, $z(0)=0.3$ are used in the figure. The irregular behavior in each component reveals that the chaotic behavior of the atmosphere can be gained from the chaoticity of the hydrosphere characteristics. 

For the downward chaos transmission from the atmosphere to the ocean, we consider the perturbed Vallis system
    \begin{equation} \label{VallisLorenzCouple}
    \begin{split}
		& \frac{du}{dt} =B\,T_d-C\,u+0.7 x,\\
		& \frac{dT_d}{dt} = u\,T_s-T_d+0.3 \cos y +0.4 y ,  \\	
		& \frac{dT_s}{dt} = -u\,T_d-T_s+1+0.5 z,
    \end{split}
    \end{equation}
where $(x,y,z)$ is the solution of the Lorenz system (\ref{LorenzSystem}) with the parameters $ \sigma = 10 $, $ r = 28 $ and $ b = 8/3 $ and the initial data $x(0)=0$, $y(0)=1$, $ z(0)=0 $. System (\ref{LorenzSystem}) possesses a chaotic attractor with these choices of the parameter values \cite{Lorenz,Sparrow}. 

\begin{figure}[H]
    	\subfigure[]{\includegraphics[width = 6.4in]{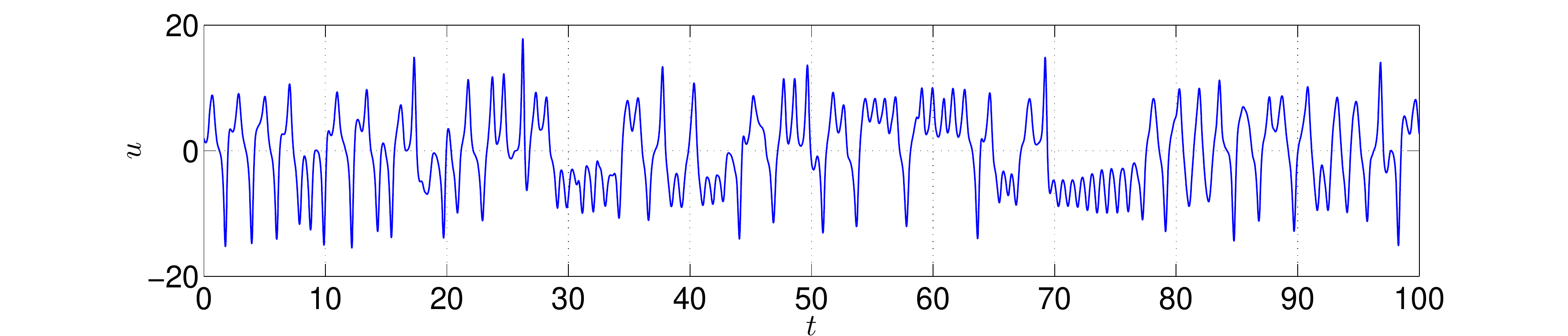}\label{VallLor-u}} \\
    	\subfigure[]{\includegraphics[width = 6.4in]{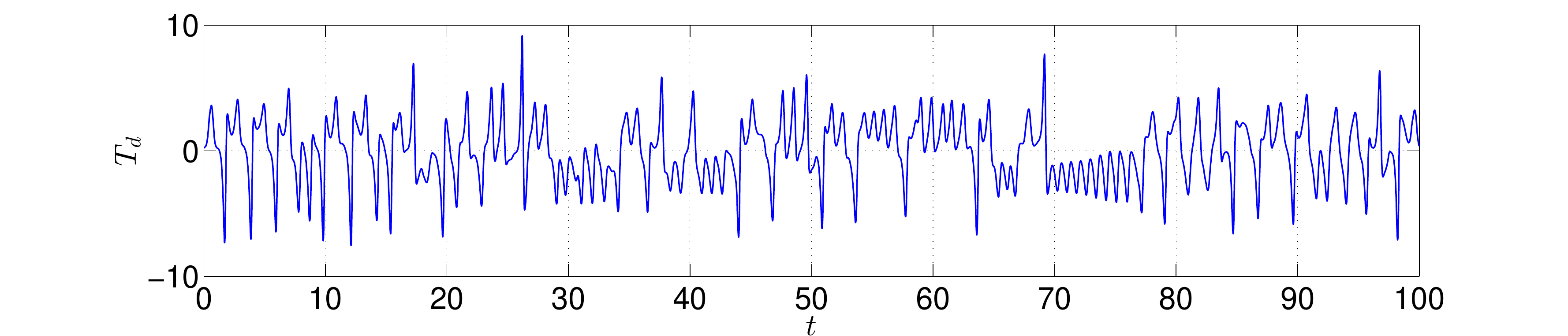}\label{VallLor-Td}}
    	\subfigure[]{\includegraphics[width = 6.4in]{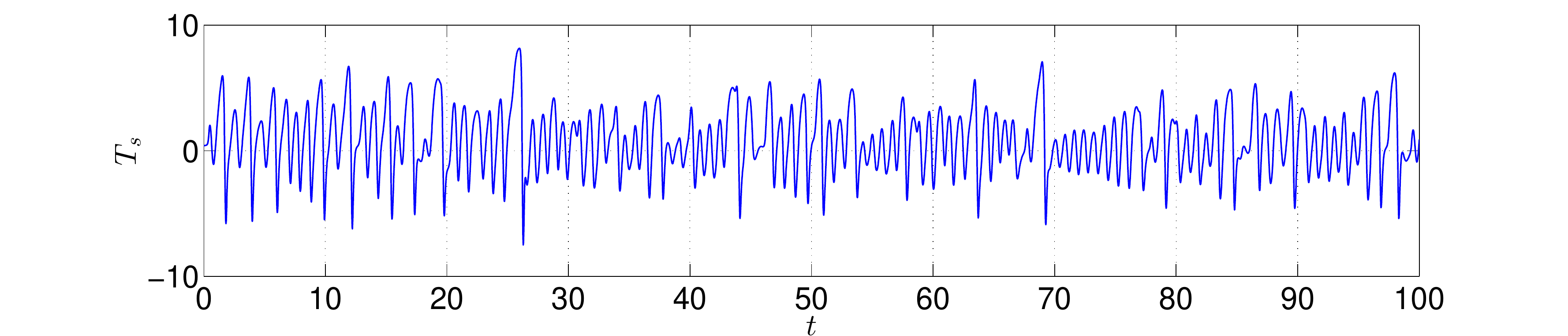}\label{VallLor-Ts}}
    	\caption{Chaotic behavior of system (\ref{VallisLorenzCouple})}
    	\label{CoupledVallisLorenz}   	
    \end{figure}

Let us take $ B = 20 $ and $ C = 7 $ in system (\ref{VallisLorenzCouple}). One can verify in this case that the corresponding unperturbed system (\ref{LorenzFormVallis}) is non-chaotic such that it possesses an asymptotically stable equilibrium. Figure \ref{CoupledVallisLorenz} depicts the solution of (\ref{VallisLorenzCouple}) with $ u(0)=2, $ $ T_d(0)=0.2 $, and $ T_s(0)=0.4 $. It is seen in Figure \ref{CoupledVallisLorenz} that the chaotic behavior of the Lorenz system (\ref{LorenzSystem}) is transmitted to (\ref{VallisLorenzCouple}). In other words, system (\ref{VallisLorenzCouple}) admits chaos even if it is initially non-chaotic in the absence of the perturbation. 
    
\section{Conclusion}
	
In this paper we discuss the possible unpredictable behavior of climate variables on a global scale. Some ENSO-like climate variabilities have a significant influence on global weather and climate. ENSO variability is suggested to be chaotic by many studies. The well-known Vallis ENSO chaotic model is one among several ENSO models that exhibit irregular behavior. The presence of chaos in ENSO can be indicated by the behavior of SST as well as ocean current velocity. We describe the dynamics of SST by the advection equation. The forcing term, based on ocean-atmosphere interaction, and the current velocity in this equation can be a source of unpredictability in SST. We prove the presence of chaos in SST dynamics by utilizing the concept of unpredictable function. The relationship and interaction between the climate variabilities, like the ones between ENSO and IOD, have attracted attention in recent literature. Constructing and understanding the dynamic models driving these phenomena are the main steps to investigate the mutual influences between these global events. The SST anomalies are closely linked to some climate variabilities teleconnections in different parts of the global ocean. We suggest that the hydrosphere characteristics can behave chaotically through the possibility of transmission of chaos between ocean neighbor subregions. We verified this transmission by different ``toy'' couples of advection equations and Vallis models. The simulations of these couples show that unpredictability can be transmitted from a local region controlled by a chaotic model into its neighbor which is described by a non-chaotic model.
		
We proposed to apply the same technique for the ``vertical'' unpredictability exchange between atmosphere and hydrosphere. In this case, the Lorenz system and the Vallis model are assigned for the atmosphere and ocean, respectively. Physically, this exchange may be done in the midst of interaction between ocean and atmosphere associated with, for example, heat exchange. By this procedure, the global unpredictability of oceanic oscillation  can be viewed as accompaniment to weather unpredictability.
		
Our approach provides a basic framework for mathematical interpretation to the irregular behavior of some global climate characteristics. It gives a way to link the local unpredictability in a component of climate system to more global scope. Further investigation can done by including different models for more climate components. Another important and interesting problem is \textit{controlling weather}. Even though the weather is too complicated to modify, a vital step can be taken toward this goal by modify the ENSO oscillation through control of chaos in its models and study the ``extension of the control'' between ENSO-like models and weather models. Chaos control in Lorenz system is still not effectively developed in the literature, where the most proposed methods are mainly depend on forcing the system into a single stable periodic behavior \cite{Chen,Yau}, and this is not adequate for real life applications. It is known that the chaos control can be achieved by using small perturbation to some parameters or variables of the system. This idea may be practically applied by making a small local artificial effect in atmosphere or hydrosphere. If we consider the positive tenor of the Lorenz's famous question, ``Does the flap of a butterfly's wing in Brazil set off a tornado in Texas?'', we can say that the small artificial climate change may prevent the occurrence or at least decrease the intensity of some extreme weather events such as cyclones, hurricanes, droughts, and floods.

	\newpage
	\appendix
	\setcounter{secnumdepth}{0}
	\section{Appendix}
	 \label{Appendix}
	 
\begin{figure}[H]
		\centering
		\includegraphics[width=0.8\linewidth]{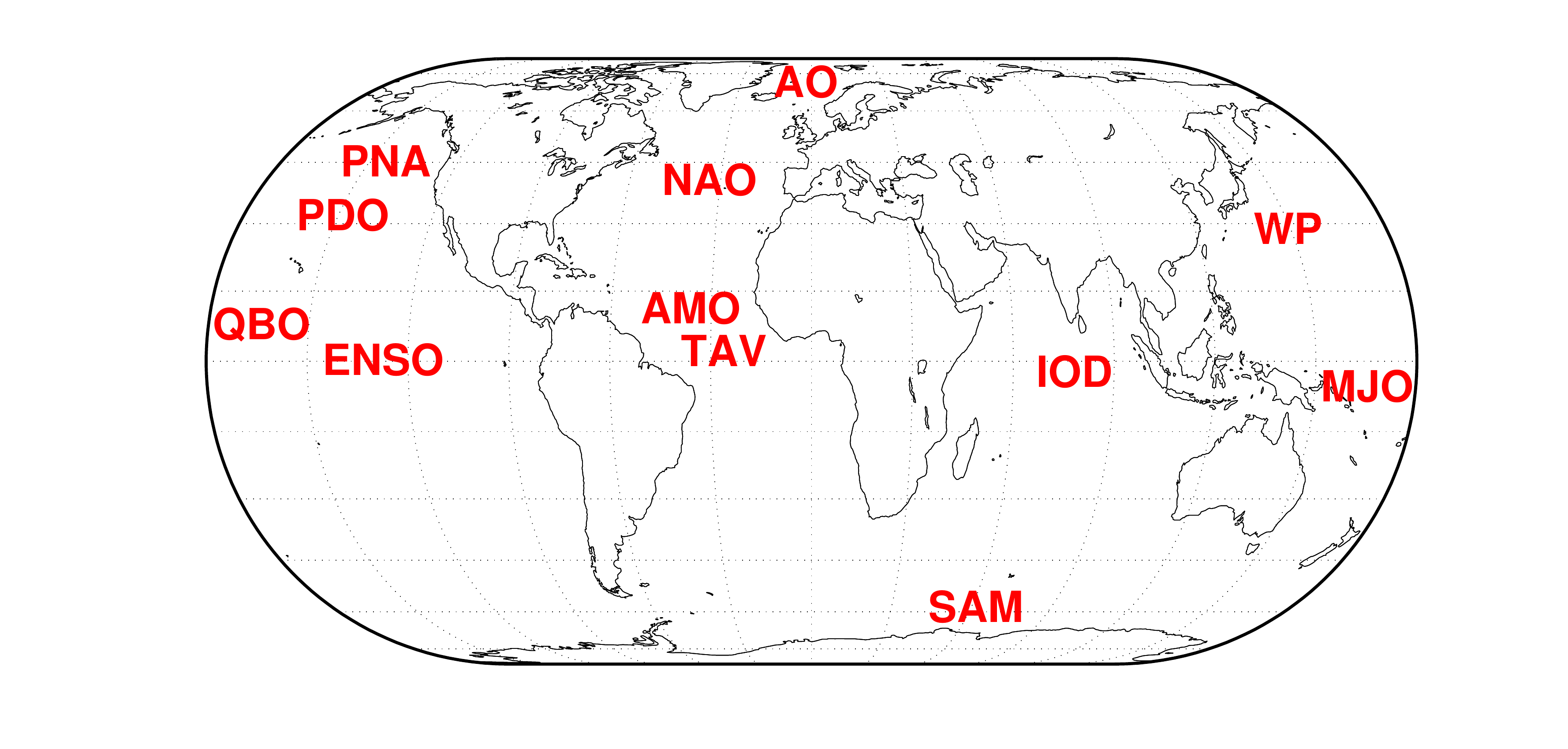}
		\vspace{-0.2cm}
		\caption{The major global climate patterns}
		\label{Wmap}	
	\end{figure}

	 In addition to ENSO and IOD, in this appendix we give a short review of the major atmospheric patterns, namely Pacific Decadal Oscillation (PDO), Atlantic Multidecadal Oscillation (AMO),  Southern Annular Mode (SAM), Tropical Atlantic Variability (TAV), North Atlantic Oscillation (NAO), Arctic Oscillation/Northern Annular Mode (AO/NAM), Madden-Julian Oscillation (MJO), Pacific/North American pattern (PNA), Quasi-Biennial Oscillation (QBO) and Western Pacific pattern (WP). Figure \ref{Wmap} shows the places of occurrence of these patterns \cite{Rosenzweig,Lehr}, and Table \ref{T1} gives brief descriptions of them \cite{Rosenzweig,Vuille,Lehr}.
	 
	\vspace{-0.2cm}
	 
	\vspace{-0.3cm}
	
    \begin{table}[H]
    	\begin{center}
    		\begin{tabular}{|M{1.5cm}|c|M{1.8cm}|M{2.2cm}|} \hline
    			Term   & Descriptions   & Main Index   & Timescale  \\ \hline
    			ENSO   &  \multicolumn{1}{m{10cm}|}{An irregularly periodical variation in sea surface temperatures over the tropical eastern Pacific Ocean}         & SST   & 3--7 years   \\\hline
    			QBO    &  \multicolumn{1}{m{10cm}|}{An oscillation of the equatorial zonal wind in the tropical stratosphere}      & SLP    & 26--30 months   \\\hline
    			PDO    &  \multicolumn{1}{m{10cm}|}{A low-frequency pattern similar to ENSO occurs primarily in the Northeast Pacific near North America}      & SST    & 20--30 years   \\\hline
    			PNA    &  \multicolumn{1}{m{10cm}|}{An atmospheric pressure pattern driven by the relationship between the warm ocean water near Hawaii and the cool one near the Aleutian Islands of Alaska}      & SLP    & 7--8 days   \\\hline
    			AO/NAM    &  \multicolumn{1}{m{10cm}|}{Defined by  westerly winds changes driven by temperature contrasts between the tropics and  northern polar areas}      & SLP    & 1--9 months   \\\hline
    			NAO    &  \multicolumn{1}{m{10cm}|}{Large scale of  pressure varies in opposite directions in the North Atlantic near Iceland in the north and the Azores in the south}      & SLP    & 9--10 days   \\\hline
    			TAV    &  \multicolumn{1}{m{10cm}|}{Like ENSO, but it exhibits a north-south low frequency oscillation of the SST gradient across the equatorial Atlantic Ocean}      & SST    & 10--15 years   \\\hline
    			AMO    &  \multicolumn{1}{m{10cm}|}{A mode of natural variability occurring in the North Atlantic Ocean and affects the SST on different modes on multidecadal timescales}      & SST    & 55--80 years   \\\hline
    			SAM    &  \multicolumn{1}{m{10cm}|}{Defined by westerly winds changes driven by temperature contrasts between the tropics and southern polar areas}      & SLP    & 30--70 days   \\\hline
    			IOD    &  \multicolumn{1}{m{10cm}|}{An irregular oscillation of sea-surface temperatures in equatorial areas of the Indian Ocean}      & SST    & 2--5 years   \\\hline
    			WP     &  \multicolumn{1}{m{10cm}|}{A low-frequency variability characterized by north-south dipolar anomalies in pressure over the Far East and western North Pacific}      & SLP    & 7--8 days   \\\hline
    			MJO    &  \multicolumn{1}{m{10cm}|}{An equatorial traveling pattern of anomalous rainfall Located in the tropical Pacific and Indian oceans}      & SLP    & 40--50 days   \\ \hline
    		\end{tabular}
    	\end{center} 
 		\vspace{-0.2cm}   	
    	\caption{The major climate variability systems}
    	\label{T1}	
    \end{table}

\end{document}